\documentclass[fleqn,usenatbib]{mnras}

\usepackage{newtxtext,newtxmath}

\usepackage[T1]{fontenc}

\DeclareRobustCommand{\VAN}[3]{#2}
\let\VANthebibliography\thebibliography
\def\thebibliography{\DeclareRobustCommand{\VAN}[3]{##3}\VANthebibliography}
\usepackage{mathtools}
\usepackage{nccmath}
\usepackage{siunitx}

\usepackage{graphicx}	
\usepackage{amsmath}	

\title[Drag force and gravitational instability --- II]{The role of the drag force in the gravitational stability of dusty planet-forming disc – II. Numerical simulations}

\author[C. Longarini et al.]{
Cristiano Longarini,$^{1,2,3,4}$\thanks{E-mail: cristiano.longarini@unimi.it}
Philip J. Armitage,$^{2,3}$
Giuseppe Lodato,$^{1}$ 
Daniel J. Price$^{4}$,
Simone Ceppi$^{1,4}$
\\
$^{1}$Dipartimento di Fisica, Università degli Studi di Milano, via Celoria 16, 20133 Milano, Italy\\
$^{2}$Department of Physics and Astronomy, Stony Brook University, Stony Brook, NY 11794, USA \\
$^{3}$Center for Computational Astrophysics, Flatiron Institute, New York, NY 10010, USA \\
$^{4}$School of Physics and Astronomy, Monash University, Clayton, VIC 3800, Australia
}

\date{Accepted XXX. Received YYY; in original form ZZZ}

\pubyear{2023}

\begin{document}
\label{firstpage}
\pagerange{\pageref{firstpage}--\pageref{lastpage}}
\maketitle

\begin{abstract}
Young protostellar discs are likely to be both self-gravitating, and to support grain growth to sizes where the particles decoupled from the gas. This combination could lead to short-wavelength fragmentation of the solid component in otherwise non-fragmenting gas discs, forming Earth-mass solid cores during the Class 0~/~I stages of Young Stellar Object evolution. We use three-dimensional smoothed particle hydrodynamics simulations of two-fluid discs, in the regime where the Stokes number of the particles ${\rm St} > 1$, to study how the formation of solid clumps depends on the disc-to-star mass ratio, the strength of gravitational instability, and the Stokes number. Gravitational instability of the simulated discs is sustained by local cooling. We find that the ability of the spiral structures to concentrate solids increases with the cooling time, and decreases with the Stokes number, while the relative dynamical temperature between gas and dust of the particles decreases with the cooling time and the disc-to-star mass ratio, and increases with the Stokes number. Dust collapse occurs in a subset of high disc mass simulations, yielding clumps whose mass is close to linear theory estimates, namely  {$1$--$10~\text{M}_\oplus$}. Our results suggest that if planet formation occurs via this mechanism, the best conditions correspond to near the end of the self-gravitating phase, when the cooling time is long and the Stokes number close to unity.
\end{abstract}




\section{Introduction}
Direct imaging of protostellar discs, in scattered light and especially at sub-mm wavelengths, has opened debate as to how the timescales for planet formation align with the established evolutionary sequence for Young Stellar Objects \citep{Adams1987}. Many discs show substructure \citep{Andrews2020} that is consistent with --- and often interpreted as --- the theoretically expected signature of planet-disc interaction \citep{Dong15,Zhang2018,Lodato2019}. A subset of these discs, including 
HL Tau \citep{HLTAU_1st}, IRS 63 \citep{irs63} and GY91 \citep{GY91} are young, with ages estimated to be less than $1 \ \text{Myr}$. Assuming a planet origin for the substructure,  a substantial part of the planet formation process must overlap with the time when protostellar discs at 10--100~au scales are likely self-gravitating. How this occurs is  unclear. Classical planetesimal-only variants of Core Accretion model \citep{CA1,Pollack1996} are not viable; meeting the timescale constraint at 10~au is already non-trivial, and at substantially larger radii sufficiently massive planet formation is not possible at all. Models with a dominant pebble component \citep[e.g.][]{Rosenthal18,Morbidelli2020,Guilera20,Chambers2021,Cummins2022,Jiang2022}, or those with large-scale core migration \citep{Levison2010}, are more promising, but the ability of these processes to form the inferred large-orbital-radius population of planets has not been fully established.

Alternatively, planets may form via gravitational instability (GI) \citep{bossGI} in a self-gravitating protostellar disk. Self-gravity is expected to be an important process during the Class 0/I phase of protostellar evolution, and is most vigorous at large orbital radii where the cooling time is short \citep{Clarke2009,Rafikov2009} and infall continues. GI in a gaseous disc, however, is a poor candidate for forming the ALMA-inferred planet population, because it substantially overshoots the inferred masses. The mass of a gaseous fragment created by  {GI} is of the order of the Jeans mass of the spiral perturbation, and for typical protoplanetary discs its value is, assuming Toomre $Q=1$ \citep{toomre,krattlod}
\begin{equation}
M_{\rm J} = \frac{4 c_{\rm s}^4}{G^2 \Sigma} \approx 12 M_{\rm Jup} \left(\frac{M_*}{1 M_\odot}\right) \left(\frac{H/R}{0.1}\right)^{3},
\end{equation}
typically between $1$--$10~\text{M}_\text{J}$ (where $M_{\rm Jup}$ is the mass of Jupiter) depending on the disc aspect ratio. However, this is the {\em initial} mass of the object: subsequently the fragment would start accreting material belonging to the accretion disc, increasing its mass and likely becoming a brown dwarf \citep{kratt1,krattlod}.


A mechanism that may allow lower mass planets to form as a consequence of GI was  proposed by \cite{fragrice}. Concentration of solid particles in spiral arms, together with vertical settling, can lead to gravitational collapse in the solid component even in a non-fragmenting gravitationally unstable gaseous disc \citep{gibbons12,gibbons14,gibbons15}. It is known that gas spiral arms act as dust traps, since they are pressure maxima \citep{fragrice,Shi16}. Solid particles collect inside them \citep{dipierrotrap}, reaching dust to gas ratio that can be of the order of unity. Conversely, the interaction between gas spiral arms and dust particles can excite them, imparting random motions that reduce the peak density and potential for collapse \citep{Riols20}. \cite{kicks} found that large dust particles experience gravitational scattering by the spiral arms, while \cite{boothclarke16} related the level of excitation of solid particles to the aerodynamical coupling and the cooling factor. Marginally coupled solid particles are less excited by spiral arms, while, in rapidly cooled discs, the level of dust excitation is higher. \cite{zhu1} confirmed this trend through 3D shearing box simulations. 

In this work, we perform three-dimensional smoothed particle hydrodynamics (SPH) simulations using the \textsc{phantom} code \citep{phantom} of gas and dust in  gravitationally unstable protostellar discs. The aim is to investigate the interplay between  {GI} and the dynamics of particles that are aerodynamically coupled to the gas, and to understand under which conditions it is possible to form solid cores inside GI spirals. In Section~\ref{s2}, we recall the basics of the aerodynamical coupling between gas and dust, and the role of dust in  {GI}. In Section~\ref{s3} we briefly describe the SPH code \textsc{phantom} \citep{phantom} and the set of simulations we have performed for this work. In Section~\ref{s4} we analyse the simulations, show our main results, and compare our findings with previous works. In Section~\ref{s5}, we discuss our findings in an evolutionary perspective. Finally, in Section~\ref{concl} we present our conclusions.

\section{Drag force and gravitational instability}\label{s2}

\subsection{Radial drift}

To zeroth order, gas dynamics is dictated by the hydrodynamical equations. Neglecting the role of the disc self gravity, in centrifugal equilibrium, the azimuthal component of the gas velocity is
\begin{equation}
    v_{\phi,g}^2 = v_k^2 + \frac{1}{\rho_g}\frac{\text{d}P}{\text{d}R},
\end{equation}
where $v_k^2 = GM_\star/R$ is the Keplerian speed and the second term is the negative contribution of the pressure gradient. It is possible to explicitly compute the pressure gradient knowing the disc temperature structure: the gas azimuthal speed is slightly sub-Keplerian, and the correction term is of the order of $(H/R)^2\sim0.01$, where $H$ is the disc thickness. In addition, the gas has also a negative radial velocity, given by viscosity.

Conversely, solid particle dynamics is dictated by both gravitational and drag force. Since dust is not supported by pressure, the basic velocity of gas and dust is different, but they interact through a drag force, that modifies the relative velocity. Writing $\Delta \mathbf{u} = \mathbf{v}_{\rm d}-\mathbf{v}_{\rm g}$, the drag force $\mathbf{F}_{\rm d}$ depends upon the relative velocity as \begin{equation}
    |\textbf{F}_{\rm d}| = \frac{1}{2} C_{\mathrm{D}} \pi s^{2} \rho \Delta u^{2},
\end{equation}
where $C_{\rm D}$ is a coefficient that depends on the drag regime, $s$ is the dust particle size and $\rho$ is the total density. The different aerodynamical coupling regimes depend upon the size of dust particles relative to the gas mean free path. It is possible to define a dimensionless parameter, called the Knudsen number \citep{mellema}, that measures this property
\begin{equation}
    \text{Kn} = \frac{9\lambda_g}{4s},
\end{equation}
where $s$ is the dust particle size and $\lambda_{\rm g} $ is the gas particles' mean free path, given by
\begin{equation}
    \lambda_{\rm g} =  \frac{5\pi}{64\sqrt{2}}\frac{\mu m_p}{\rho_\text{g}\sigma_\text{coll}},
\end{equation}
where $\sigma_\text{coll}=2.367\times10^{-15}\text{cm}^2$ is the cross-section of gas particles, $\mu = 2.1$ is the mean molecular weight, $\rho_{\rm g}$ is the gas density and $m_p$ is the mass of the proton. Typically, in protoplanetary discs, $\text{Kn}>1$, meaning that the dust particles' size is smaller than the gas mean free path: this is called the Epstein regime. However, when the disc is very massive, there is a transition between Epstein and Stokes regime ($\text{Kn}<1)$.  {The main impact of this transition is that in the inner denser region of the disc, where particles’ size is comparable with the gas mean free path, the Stokes number is higher compared to Epstein regime}. The $C_\text{D}$ coefficient is given by \citep{Fassio70}

\begin{equation}
    C_{\mathrm{D}}=\left\{\begin{array}{ccc}
\frac{8}{3} \frac{c_{g}}{\Delta u} & \text{Kn}>1 & \\
24 \text{Re}^{-1} & \text{Kn}<1, & \text{Re}<1 \\
24 \text{Re}^{-0.6} & \text{Kn}<1, & 1<\text{Re}<800 \\
0.44 & \text{Kn}<1, &  \text{Re}>800
\end{array}\right.
\end{equation}
where $c_g$ is the gas sound speed and the Reynolds number based upon the difference of velocity is given by
\begin{equation}
    \text{Re} = \frac{2s\Delta u}{\nu} = 4\frac{s\Delta u}{\lambda_g c_g},
\end{equation}
and the last equivalence is true for collisional viscosity. We also define the stopping time, i.e. the time needed to modify the relative velocity between gas and dust. The longer it is, the less the particles are coupled. For spherical grains, it is given by
\begin{equation}
    t_\text{s} = \frac{m_d\Delta u}{|\mathbf{F}_D|} =  \frac{4\pi\rho_0 s^3 \Delta u}{3|\mathbf{F}_D|} = \frac{8\rho_0s}{3C_D\rho\Delta u},
\end{equation}
where $m_d$ is the dust grain mass and $\rho_0$ is the dust grain's intrinsic density. In order to measure the strength of the aerodynamical coupling, we define a  dimensionless Stokes number as the ratio between the stopping time and the dynamical one: in general, its expression is
\begin{equation}
    \text{St} = t_\text{s} \Omega = \frac{8}{3 C_{\mathrm{D}}}\left(\frac{\rho_{\mathrm{0}}}{\rho}\right)\left(\frac{v_{\mathrm{k}}}{\Delta u}\right)\left(\frac{s}{r}\right),
\end{equation}
and, in the Epstein regime, its value is
\begin{equation}
    \text{St} = 1 \left(\frac{\Sigma}{0.2\text{g}/\text{cm}^2}\right)^{-1}\left(\frac{\rho_0}{3\text{g}/\text{cm}^3}\right) \left(\frac{s}{1\text{mm}}\right) .
\end{equation}
The smaller the Stokes number is, the more tightly the particles are coupled: so, for $\text{St}\to0$, dust dynamics follows the gas dynamics, while for $\text{St}\to \infty$, the two fluids do not influence each other aerodynamically. In a smooth axisymmetric disc, the drag force has dramatic physical consequences. Because of the azimuthal difference of speed between gas and dust, solid particles experience a headwind that slows them down, giving them a negative radial velocity. This effect is called radial drift, and its timescale for ${\rm St} \sim 1$ is of the order of $\sim 100$yr, several orders of magnitude shorter than the disc lifetime. This is the ``metre sized barrier", so called because ${\rm St} \sim 1$ corresponds to roughly this physical size in the inner disc \citep{metresized}. Radial drift can be stopped in discs where the presence of a gas pressure maximum creates a ``dust trap''. Indeed, in a gas pressure maximum, the pressure gradient is zero, and hence the velocity difference between gas and dust vanishes and no radial drift occurs. Several mechanisms have been proposed to form dust traps, such as gaps made by planets \citep{dustgap1,mellema,dustgap2,pinilladust}, zonal flows \citep{Johansen09,Simon14}, or spirals induced by gravitational instabilities \citep{dipierrotrap}. Concentration of solid material by aerodynamic effects must be considered in assessing models for planet formation 
in early protoplanetary stages. 

\subsection{Gravitational instability}

In an axisymmetric thin disc composed of a single fluid, the dispersion relation for linear, tightly-wound density waves is
\begin{equation}
    D(\omega, k, m) = (\omega-m\Omega)^2 -c^2k^2+2\pi G \Sigma {|k|} -\kappa^2,
\end{equation}
where $\omega$ is the perturbation frequency, $\Omega(R)$ is the angular velocity, $m/R$ is the azimuthal wavenumber of the perturbation, $c(R)$ the sound speed of the fluid, $k$ the perturbation radial wavenumber, $\kappa(R)$ the epicyclic frequency and $\Sigma(R)$ the disc surface density. The well-known instability criterion for axisymmetric disturbances is \citep{toomre}
\begin{equation}
    Q = \frac{c\kappa}{\pi G \Sigma} <1,
\end{equation}
that identifies the parameter space for which $\omega^2<0$. It is possible to generalize the instability taking into account a second fluid component, i.e. particles in the protostellar case. In the context of galactic dynamics, the gravitational role of a second component has been investigated by \cite{jogsolo} and \cite{bertinromeo}. \cite{longa2fl} applied this method to protoplanetary discs, taking into account also the role of the drag force between gas and dust. When the second component is considered, the outcome of  {GI} can significantly change. In the protostellar scenario, the three fundamental parameters are the relative concentration of the dust component to that of the gas
\begin{equation}
    \epsilon = \frac{\Sigma_d}{\Sigma_g},
\end{equation}
the dust relative temperature 
\begin{equation}
    \xi = \left(\frac{c_d}{c_g}\right)^2,
\end{equation}
where $c_d$ is the dust dispersion velocity, and the Stokes number 
\begin{equation}
    \text{St} = t_s\Omega.
\end{equation}
Two regimes of instability can be identified, depending on whether the instability is triggered by the gas or the dust. When the instability is driven by the gas component, the most unstable wavelength is close to the gas-only one, and the role of dust is negligible. Conversely, when the instability is controlled by the dust, the perturbation wavelength is much smaller. This different kind of instability happens when the dust is sufficiently cold, decoupled and abundant: the threshold between these two regimes is given by \citep{bertinromeo}
\begin{equation}\label{dustdr}
    \epsilon \xi^{-1/2}>1,
\end{equation}
or, taking into account the drag interaction \citep{longa2fl}
\begin{equation}\label{dustdrdrag}
    \epsilon \xi^{-1/2}(1+0.72\text{St}^{-1.36})>1.
\end{equation}
Dust driven instability could have important consequences for planet formation because when the most unstable wavelength (i.e. the Jeans wavelength) is smaller, the Jeans mass will be small as well. In the gas-only model, the Jeans length is
\begin{equation}
    \lambda^\text{1f}_J = \frac{2c_g^2}{G\Sigma_g} = \sqrt{\frac{2}{\pi}} \frac{c_g\Omega}{G\rho_g},
\end{equation}
where we used $\rho_g = \Sigma_g/\sqrt{2\pi}H$, and the Jeans mass is 
\begin{equation}
    M^\text{1f}_J = \frac{4}{3}\pi \rho_g \left(\lambda^\text{1f}_J\right)^3 = \frac{4\pi}{3}\left(\frac{2}{\pi}\right)^{3/2} \frac{c_g^3\Omega^3}{G^3\rho_g^2}.
\end{equation}
When the instability becomes dust driven, the value of the most unstable wavelength changes dramatically. It is possible to compute it through the dispersion relation of the two fluid component model with drag force \citep{longa2fl}, and its value depends on the Stokes number, the relative temperature and the dust to gas ratio. In particular, the gas-only limit is obtained for $\text{St}\to0$, $\epsilon\to0$ and $\xi\to1$. For convenience, we define the ratio between the Jeans wavelength of one fluid component model and of the two fluid one with drag force as $\Lambda(\epsilon,\xi,\text{St})$: the value of $\Lambda$ is between 0 and 1, where $\Lambda=1$ is the one fluid limit. Indeed, by computing the most unstable wavelength, it is possible to extract $\Lambda$.  Hence, the Jeans mass of the two-component fluid model is
\begin{equation}\label{jmass_complete}
    M^\text{2f}_J = M^\text{1f}_J \Lambda^3.
\end{equation}
In Appendix~\ref{applambda}, we show plots of $\Lambda^3$ as a function of $(\epsilon,\xi,\text{St})$.

\section{Numerical simulations}\label{s3}
In this work, we perform numerical SPH simulations of gas and dust protostellar discs using the code \textsc{phantom} \citep{phantom}. This code is widely used in the astrophysical community to study gas and dust dynamics in accretion discs \citep{enriphantom,simophantom}, both in a single fluid mixture \citep{benniphantom} or dust-as-particles approach \citep{alyphantom}. In this work, we use the dust-as-particles formulation.

\subsection{Two-fluid gas and dust mixtures}
The dust formulation is based on the continuum fluid equations in the form
\begin{equation}
    \frac{\partial \rho_{\mathrm{g}}}{\partial t}+\left(\mathbf{v}_{\mathrm{g}} \cdot \nabla\right) \rho_{\mathrm{g}} =-\rho_{\mathrm{g}}\left(\nabla \cdot \mathbf{v}_{\mathrm{g}}\right),
\end{equation}

\begin{equation}
    \frac{\partial \rho_{\mathrm{d}}}{\partial t}+\left(\mathbf{v}_{\mathrm{d}} \cdot \nabla\right) \rho_{\mathrm{d}} =-\rho_{\mathrm{d}}\left(\nabla \cdot \mathbf{v}_{\mathrm{d}}\right),
\end{equation}

\begin{equation}
    \frac{\partial \mathbf{v}_{\mathrm{g}}}{\partial t}+\left(\mathbf{v}_{\mathrm{g}} \cdot \nabla\right) \mathbf{v}_{\mathrm{g}} =-\frac{\nabla P}{\rho_{\mathrm{g}}}+\frac{K}{\rho_{\mathrm{g}}}\left(\mathbf{v}_{\mathrm{d}}-\mathbf{v}_{\mathrm{g}}\right),
\end{equation}

\begin{equation}
    \frac{\partial \mathbf{v}_{\mathrm{d}}}{\partial t}+\left(\mathbf{v}_{\mathrm{d}} \cdot \nabla\right) \mathbf{v}_{\mathrm{d}} =-\frac{K}{\rho_{\mathrm{d}}}\left(\mathbf{v}_{\mathrm{d}}-\mathbf{v}_{\mathrm{g}}\right),
\end{equation}
where the subscripts $g$ and $d$ refer to gas and dust properties. We define the stopping time, that is given by
\begin{equation}\label{tstop}
    t_s \equiv \frac{\rho_\text{g}\rho_\text{d}}{K(\rho_\text{g}+\rho_\text{d})},
\end{equation}
where the drag coefficient $K$ depends on the physical drag regime the system is in. In general, it is given by
\begin{equation}
    K=\rho_{\mathrm{g}} \rho_{\mathrm{d}} \frac{1}{2} C_{\mathrm{D}} \frac{\pi s^{2}}{m_{\text {d }}}|\Delta v|,
\end{equation}
where $C_D$ is defined as before.

In the \textsc{phantom} implementation, the two phases are modelled as two distinct sets of particles: hereafter, we adopt the convention from \citet{monaghan2fl}, and refer to gas particles with the subscripts $a,b,c$ and to dust particles with $i,j,k$. Gas and dust densities are computed by weighted summation over the particles of the same type according to
\begin{equation}
    \rho_{a}=\sum_{b} m_{b} W_{a b}\left(h_{a}\right) ; \quad h_{a}=h_{\mathrm{fact}}\left(\frac{m_{a}}{\rho_{a}}\right)^{1 / 3},
\end{equation}

\begin{equation}
    \rho_{i}=\sum_{j} m_{j} W_{i j}\left(h_{i}\right) ; \quad h_{i}=h_{\mathrm{fact}}\left(\frac{m_{i}}{\rho_{i}}\right)^{1 / 3},
\end{equation}
where the kernel $W$ is the same for gas and dust, $h$ is the smoothing length and $h_\text{fact} = 1/2$. The drag terms of the equations of motion are discretized 
by using a 
``double hump" kernel \citep{DHkernel} that, instead of the bell-shaped kernel, goes to zero at $r=0$ and has a peak at $r/h\lesssim1$. \footnote{\cite{laibeprice1} showed that using a double-hump kernel gives a factor of 10 better accuracy at no additional computational cost. For further information, see Section $\mathbf{2.13.4}$ of \cite{phantom}.} In these simulations, we are using the velocity reconstruction procedure presented in \cite{reconstruction}.

\subsection{Heating and cooling}
In order to take into account the effect of cooling or heating phenomena, we write the complete equation for the evolution of gas internal energy $e$
\begin{equation}
    \frac{\partial e}{\partial t}+\left(\mathbf{v}_{\mathrm{g}} \cdot \nabla\right) e =-\frac{P}{\rho_{\mathrm{g}}}\left(\nabla \cdot \mathbf{v}_{\mathrm{g}}\right)+\Lambda_\text{shock} - \frac{\Lambda_\text{cool}}{\rho_g}+\frac{\Lambda_{\mathrm{drag}}}{\rho_g},
\end{equation}
where the first term on the RHS is the $P\text{d}V$ work, the second is a heating term due to the shock viscosity, the third is the cooling of the disc and the last term is the drag heating term. In this work, we assume an adiabatic equation of state. For an ideal gas, it is possible to link pressure and density as follows
\begin{equation}
    P = (\gamma-1)\rho_g e = \frac{c_g^2\rho_g}{\gamma},
\end{equation}
where $\gamma = 5/3$ and $c_g$ is the gas sound speed, that is initialized as a power law $c_g\propto R^{-0.25}$. 

The shock viscosity term can be written as
\begin{equation}
    \Lambda_\text{shock} = c_1 \alpha^\text{AV} \frac{h_g}{H} + c_2 \beta^\text{AV}\left(\frac{h_g}{H}\right)^2,
\end{equation}
where $c_1,c_2$ are two numerical factors, whose dimension is a specific energy per unit time, $\alpha^\text{AV}$ and $\beta^\text{AV}$ are respectively the linear and the quadratic viscosity coefficients, and $h_g/H_g$ is related to the numerical resolution\footnote{This quantity tells how many smoothing lengths are included in the disc thickness.}. The viscosity term is dissipative, so it heats the disc. In the \textsc{phantom} simulations we are not using the \textsc{disc-viscosity} flag, meaning that the shock capturing viscosity is not described by an $\alpha_\text{SS}$ \citep{shakurasun} prescription. We did so since in these systems the main driver of angular momentum transport is GI.

For the cooling we use the prescription from \cite{Gammiecool} and \cite{rice04}, in which the cooling time $t_\text{cool}$ is proportional to the dynamical time, with a factor of proportionality $\beta_\text{cool}$
\begin{equation}
    t_\text{cool} = \beta_\text{cool}\Omega^{-1}
\end{equation}
Under the assumption that
the transfer of angular momentum driven by gravito-turbulence occurs locally \citep{Lodato04,Bethune21}, we can relate the cooling parameter to an effective $\alpha-$viscosity parameter
\begin{equation}\label{alphacool}
    \alpha_\text{GI} = \frac{4}{9}\frac{1}{\gamma(\gamma-1)\beta_\text{cool}}.
\end{equation}
We need to introduce a cooling prescription for the system in order to trigger  {GI}. Finally, the drag heating term is
\begin{equation}
    \Lambda_\text{drag} = K|\mathbf{v}_d-\mathbf{v}_g|^2.
\end{equation}
Currently, \textsc{phantom} neglects any thermal coupling between the dust and the gas, aside from the drag heating.

\subsection{Numerical setup}

We perform simulations with three different disc-to-star mass ratios $M_d/M_\star = \{0.05,0.1,0.2\}$, three different cooling times $\beta_\text{cool}=\{8,10,15\}$ and two different dust particle sizes, for a total of 18 simulations. See below for further details. We first initialize a gas-only disc around a solar mass star, with $R_\text{in} = 0.25$au, $R_\text{out} = 25$au, and $\Sigma_g\propto R^{-1}$. We set the aspect ratio so that $Q_\text{ext}=2$ initially and because of the cooling, it decreases, eventually reaching $Q=1$. The shock viscosity coefficients $\alpha^\text{AV} = 0.1$, $\beta^\text{AV} = 0.2$ as in \cite{fragrice}. As a test, we performed a simulation with $\beta^\text{AV} = 2$, and we did not find any differences in terms of the relevant quantities in this work. The self-gravity of both gas and dust is taken into account, as well as the dust back-reaction. We performed simulations at two different numerical resolutions: the standard runs are performed with $N_g = 10^6$ and the  high resolution ones with $N_g = 2\times 10^6$. In both cases, $N_d = N_g/5$, where $N_g$ and $N_d$ are the number of gas and dust particles respectively. We verify that the results are consistent with the different resolutions.

We let the system evolve for an outer thermal time ($\beta_\text{cool}\Omega^{-1}$ at the outer radius), and then we add dust particles and evolve for a further 5 outer dynamical times (i.e. $5\times10^3$ inner dynamical times). {The dust particles are added proportional to the gas distribution.}\footnote{We benchmarked our simulations with the ones of \cite{rice04}, as they started with an initial uniform dust distribution, with a fixed thickness. We obtained the same results, since dust trapping is very efficient in these systems. {Something that should be pointed out is that \cite{rice04} did not account for self-gravity acting on the solid particles.}} {Distributing dust particles proportional to the gas is a valid assumption only for St <10. \cite{Shi16} and \cite{zhu1} showed that for uncoupled particles gravitational interactions and stirring become quite relevant, and hence dust distribution is closer to a uniform one. This is particularly clear for $\text{St}\sim100$, where. In our simulations of large dust grains, although the initial distribution is proportional to the gas one, we observe that the spiral is less prominent compared to smaller grains.}

Dust back-reaction and dust self-gravity are always taken into account. Since the aim of the work is to study the effect of the drag force in  {GI}, we use two different dust sizes: a larger one, to reproduce weakly coupled solid particles, and a smaller one, to study  {marginally} coupled particles. Since the disc mass is different across our set of simulation, we chose to adapt the particles' size in order to obtain the same Stokes number distribution. To do so, we computed the radially averaged Stokes number as a function of the particle size for different disc-to-star mass ratio, starting from the initial conditions of the gas disc, taking into account the transition between Epstein and Stokes regime. We decided to choose the small particles' size so that the radially averaged Stokes number $\langle\text{St}\rangle=8$, and for the large ones $\langle\text{St}\rangle=40$: in this way we effectively cover a Stokes number range from 1 to 10 with smaller grains, and from 10 to 100 for larger ones. One exception is that the small dust particles for the highest disc-to-star mass ratio simulations are chosen to have $\langle \text{St}\rangle = 16$ for computational reasons. In addition,  as we will show in the next section, a higher disc-to-star mass ratio makes the dust unstable, and collapse can happen: for that reason, dust in simulations \textbf{S16,\ S17} and \textbf {S18} is evolved only for an outer dynamical time. In every simulation, dust intrinsic density is fixed $\rho_0 = \SI{5}{g/cm^3}$.

\begin{figure*}
\includegraphics{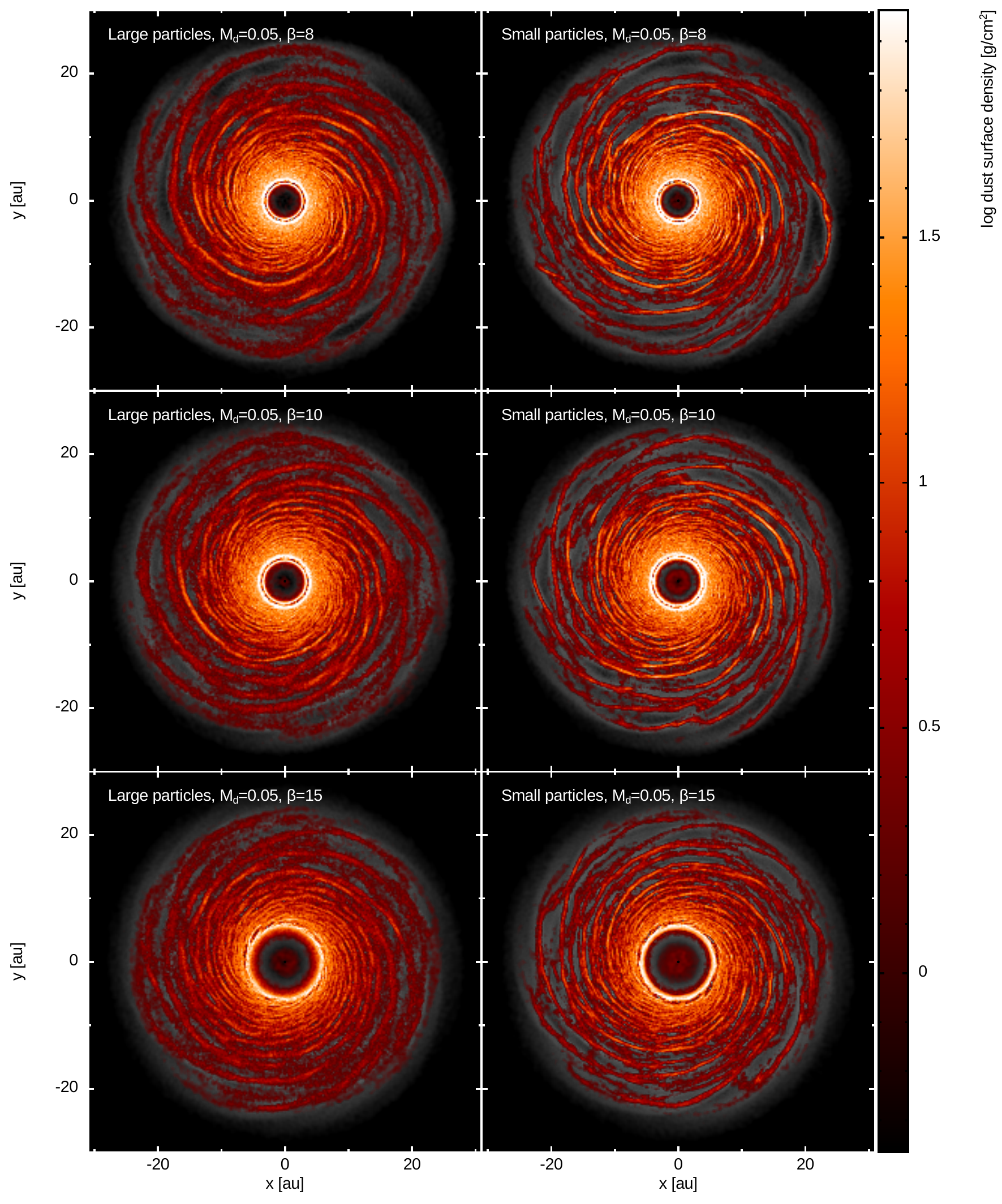}
    \caption{\textit{Dust dynamics in gas spiral arms}: large and small dust particles surface density for different cooling factor and $M_d/M_\star=0.05$.}
        \label{snapshotz1}
\end{figure*}
\begin{figure*}
\includegraphics{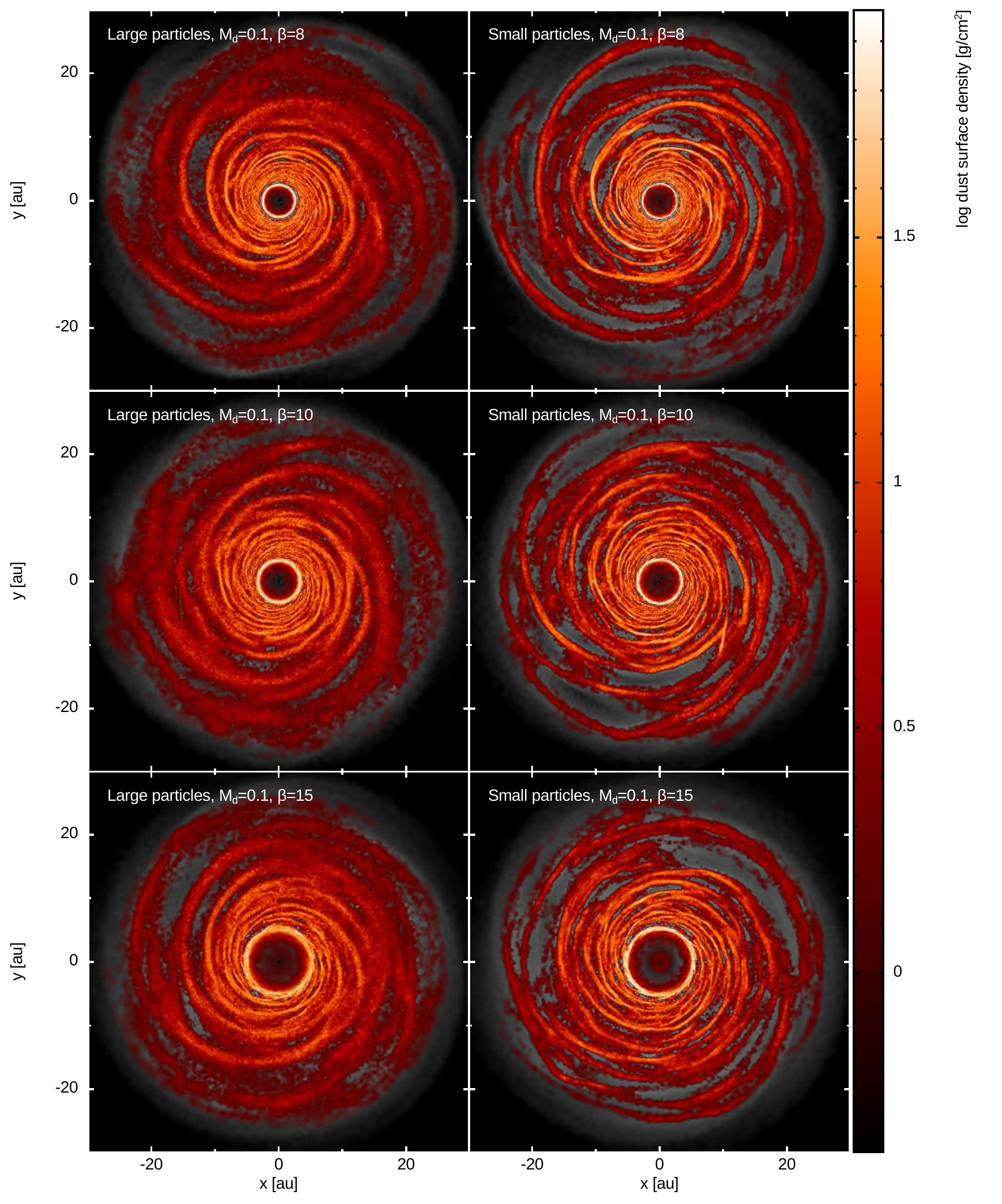}
    \caption{\textit{Dust dynamics in gas spiral arms}: large and small dust particles surface density for different cooling factor and $M_d/M_\star=0.1$.}
    \label{snapshotz2}
\end{figure*}
\begin{figure*}
\includegraphics{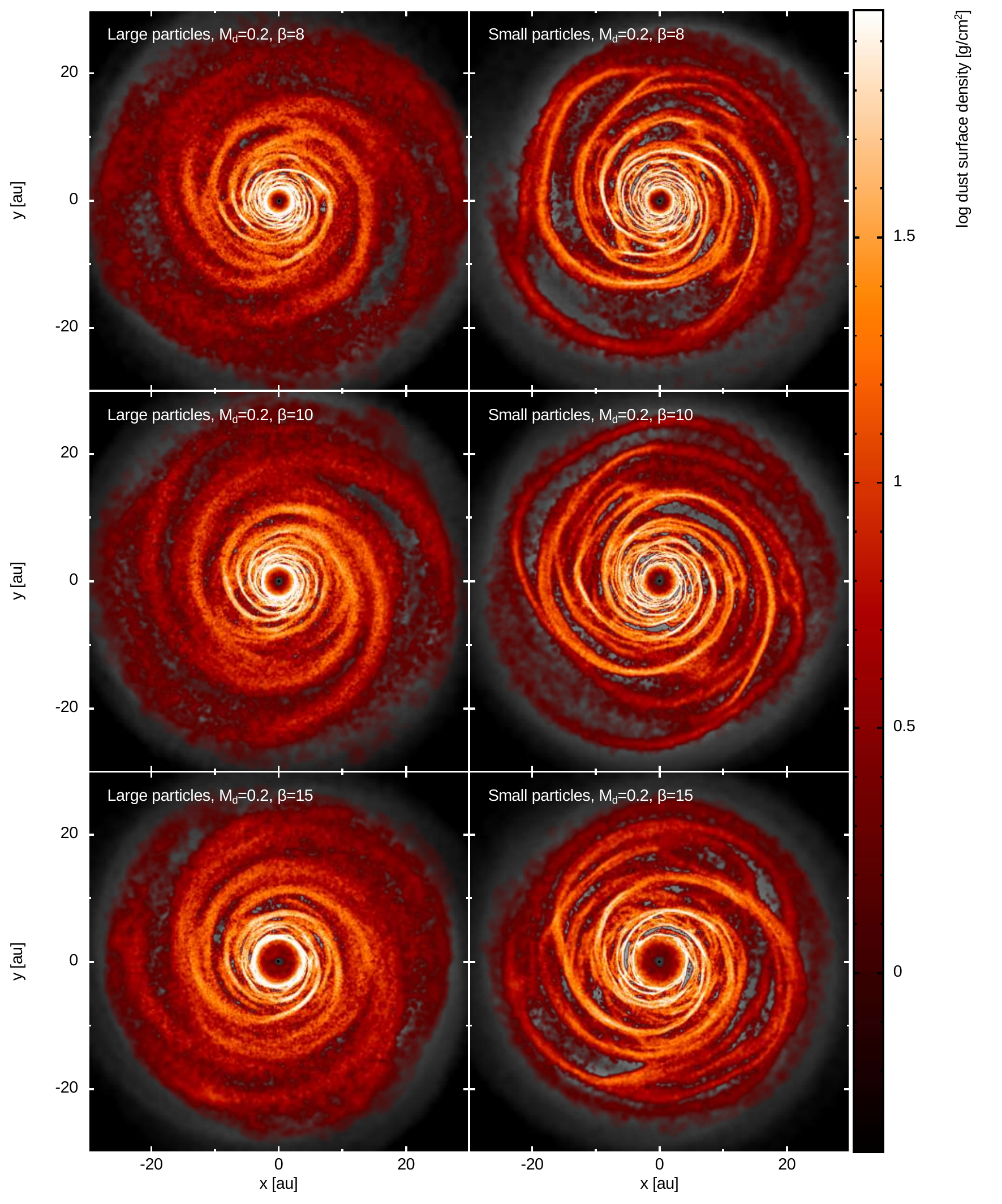}
    \caption{\textit{Dust dynamics in gas spiral arms}: large and small dust particles surface density for different cooling factor and $M_d/M_\star=0.2$.}
    \label{snapshotz3}
\end{figure*}

Self-gravitating discs may be more radially extended than our models, which can be rescaled. If we rescale the outer radius of a factor $\lambda$, how does the dust particles' size need to be rescaled? Since $\text{St}\propto{s}/{\Sigma}\propto sR_\text{out}^2$, if we change the outer radius according to $R_\text{out}' = \lambda R_\text{out}$, in order to maintain the same Stokes number, the corresponding rescaling for dust particles size should be $s' = \lambda^{-2}s$. Hence, if we consider a larger disc with $R_\text{out}' = 10R_\text{out}=250$~au, the corresponding particle sizes should be rescaled as $s' = s/100$. Dust properties are summarized in Table~\ref{params456}, where we also include the rescaled dust particles' size. Snapshots of the hydrodynamical simulations are shown in Figures~\ref{snapshotz1}, \ref{snapshotz2}, \ref{snapshotz3}.

\begin{table}\caption{Parameters of simulations: disc-to-star mass ratio $M_d/M_\star$, cooling factor $\beta_\text{cool}$, size of dust particles $s$, average Stokes number $\langle\text{St}\rangle$ and corresponding dust particles size in a 10 times bigger disc $s_{10}$.}
\centering
\begin{tabular}{llllll}
\textbf{Simulation} & {$M_d/M_\star$} & {$\beta_\text{cool}$} & {s [cm]} & $\langle\text{St}\rangle$ & s$_{10}$ [cm] \\ \cline{1-6} 
\cr
\textbf{S1} & 0.05 & 8 & 300 & 40 & 3 \\
\textbf{S2} & 0.05 & 10 & 300 & 40 & 3 \\
\textbf{S3} & 0.05 & 15 & 300 & 40 & 3 \\
\textbf{S4} & 0.05 & 8 & 60 & 8 & 0.6 \\
\textbf{S5} & 0.05 & 10 & 60 & 8 & 0.6 \\
\textbf{S6} & 0.05 & 15 & 60 & 8 & 0.6 \\
\textbf{S7} & 0.1 & 8 & 600 & 40 & 6 \\
\textbf{S8} & 0.1 & 10 & 600 & 40 & 6 \\
\textbf{S9} & 0.1 & 15 & 600 & 40 & 6 \\
\textbf{S10} & 0.1 & 8 & 120 & 8 & 1.2 \\
\textbf{S11} & 0.1 & 10 & 120 & 8 & 1.2 \\
\textbf{S12} & 0.1 & 15 & 120 & 8 & 1.2 \\
\textbf{S13} & 0.2 & 8 & 1500 & 40 & 15 \\
\textbf{S14} & 0.2 & 10 & 1500 & 40 & 15 \\
\textbf{S15} & 0.2 & 15 & 1500 & 40 & 15 \\
\textbf{S16} & 0.2 & 8 & 600 & 16 & 6 \\
\textbf{S17} & 0.2 & 10 & 600 & 16 & 6 \\
\textbf{S18} & 0.2 & 15 & 600 & 16 & 6
\end{tabular}\label{params456}
\end{table}


\section{Analysis and Results}\label{s4}
In this section, we present the analysis of the numerical simulations. Since the simulations use a two-fluid approach \citep{laibeprice1,laibeprice2}, gas and dust particles are treated as two different sets of particles, thus they occupy different locations and carry their own physical information. In order to obtain properties that depend on both gas and dust, such as dust to gas ratio, or Stokes number, we interpolate gas properties to the location of dust particles.  In addition, since dust is modelled as a pressureless fluid, it has no internal energy and no thermal sound speed. However, stirring phenomena induce a velocity dispersion onto dust particles \citep{youdinstirr}: to obtain this quantity, we compute it with an SPH interpolation over neighbouring dust particles, via
\begin{equation}
    c_{d,i}^2 =\sum_{j=1}^{N_{\text {neigh }}} m_{j} \frac{\left(v_{d,i}-v_{d, j}\right)^{2}}{\rho_{j}} W_{ij}\left( h_i\right).
\end{equation}
For our analysis, we will mainly focus our attention on the dust to gas ratio $\epsilon$, the relative temperature between gas and dust $\xi=(c_d/c_g)^2$, the Stokes number $\text{St}$, the cooling factor $\beta_\text{cool}$ and the disc-to-star mass ratio $M_d/M_\star$. These parameters are related to different physical phenomena: the  {dust to gas} ratio and the relative temperature trace dust trapping and dust excitation respectively, the cooling factor is linked to the gas spiral amplitude (and hence to the strength of GI), the Stokes number determines the power of the aerodynamical coupling and the disc-to-star mass ratio is connected to the spiral morphology. 

\subsection{Dust trapping: the $\epsilon$ parameter}

\begin{figure}
	\includegraphics[scale=0.475]{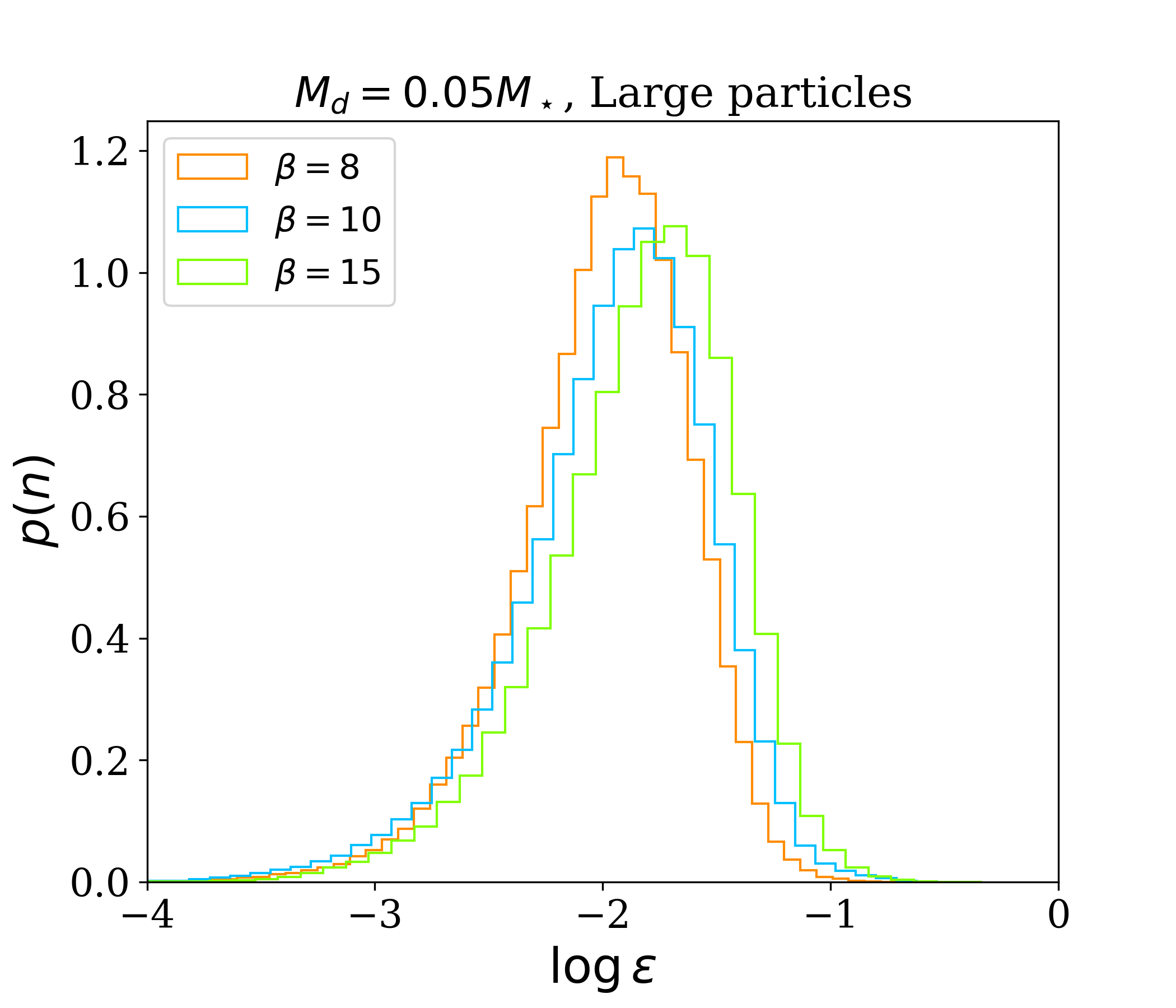}
	\includegraphics[scale=0.475]{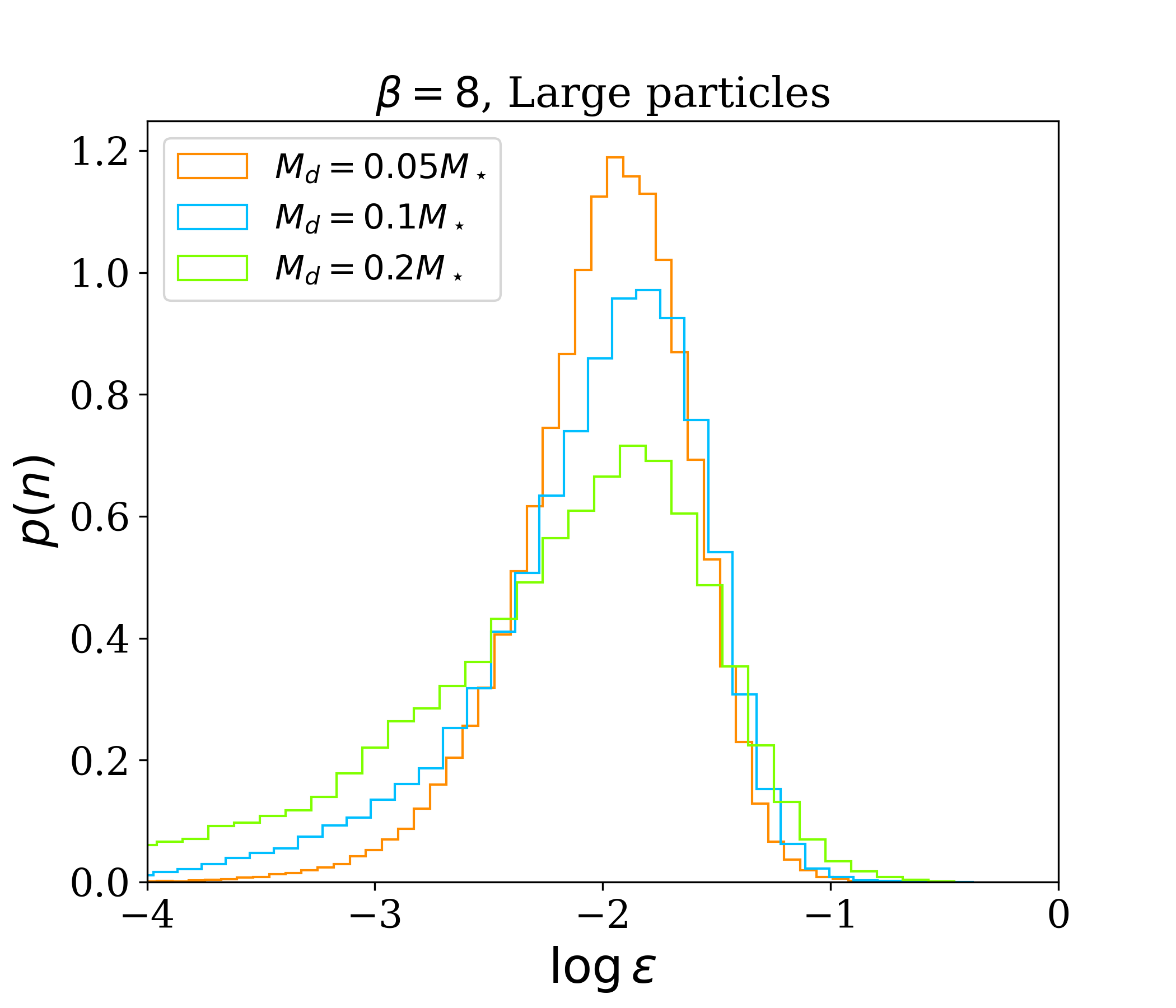}
    \caption{Distribution of the dust to gas ratio for different values of cooling factor (top panel) and disc to star mass ratio (bottom panel). The simulations shown in these plots are \textbf{S1,S2,S3,S7,S13}.}
    \label{epsilon_histo}
\end{figure}

\begin{figure}
	\includegraphics[scale=0.475]{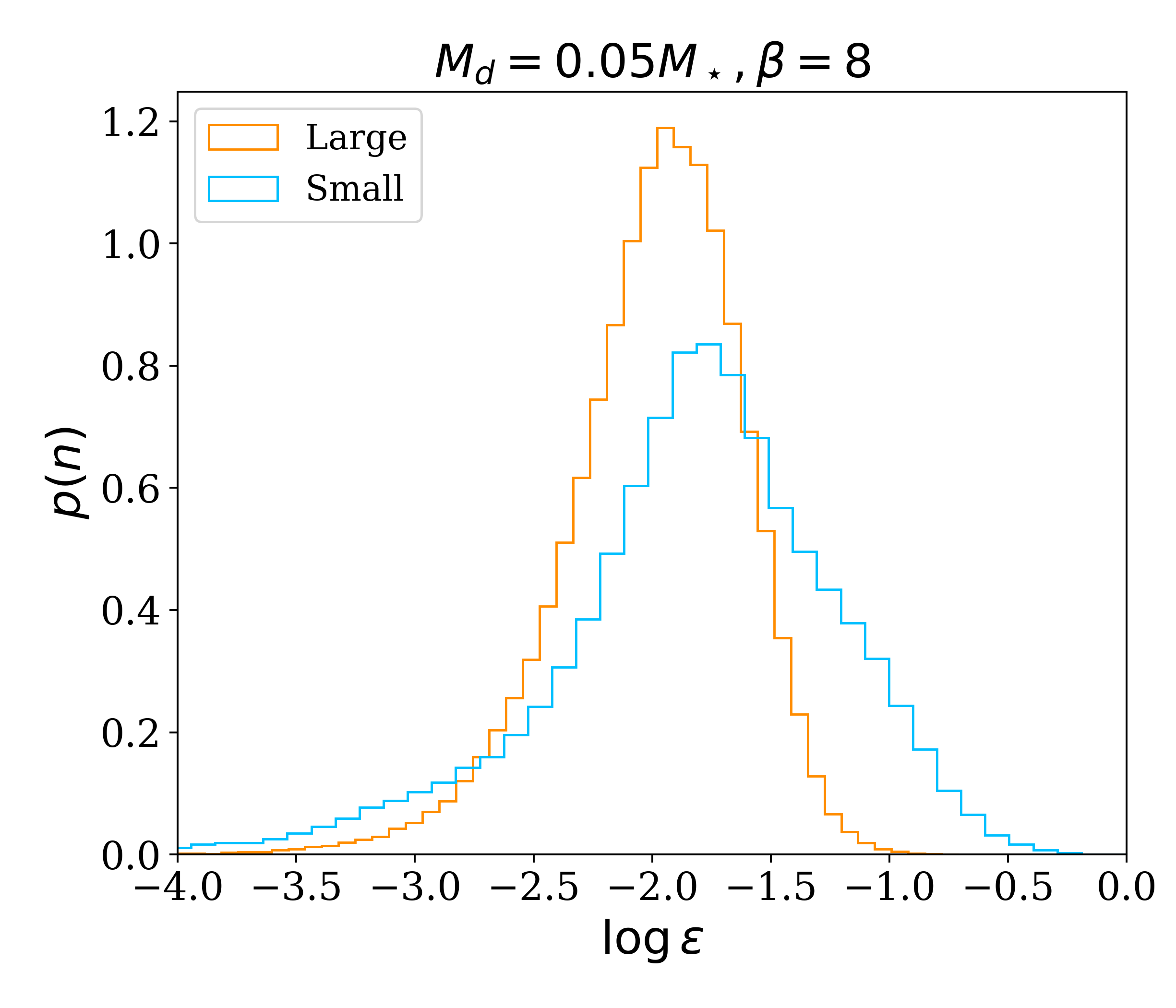}
    \caption{Comparison of the distribution of dust to gas ratio of large (orange line) and small (blue line) dust particles. The simulations shown in this plot are \textbf{S1} and \textbf{S4}.}
    \label{epsilon_largesmall}
\end{figure}

\begin{figure}
	\includegraphics[scale=0.41]{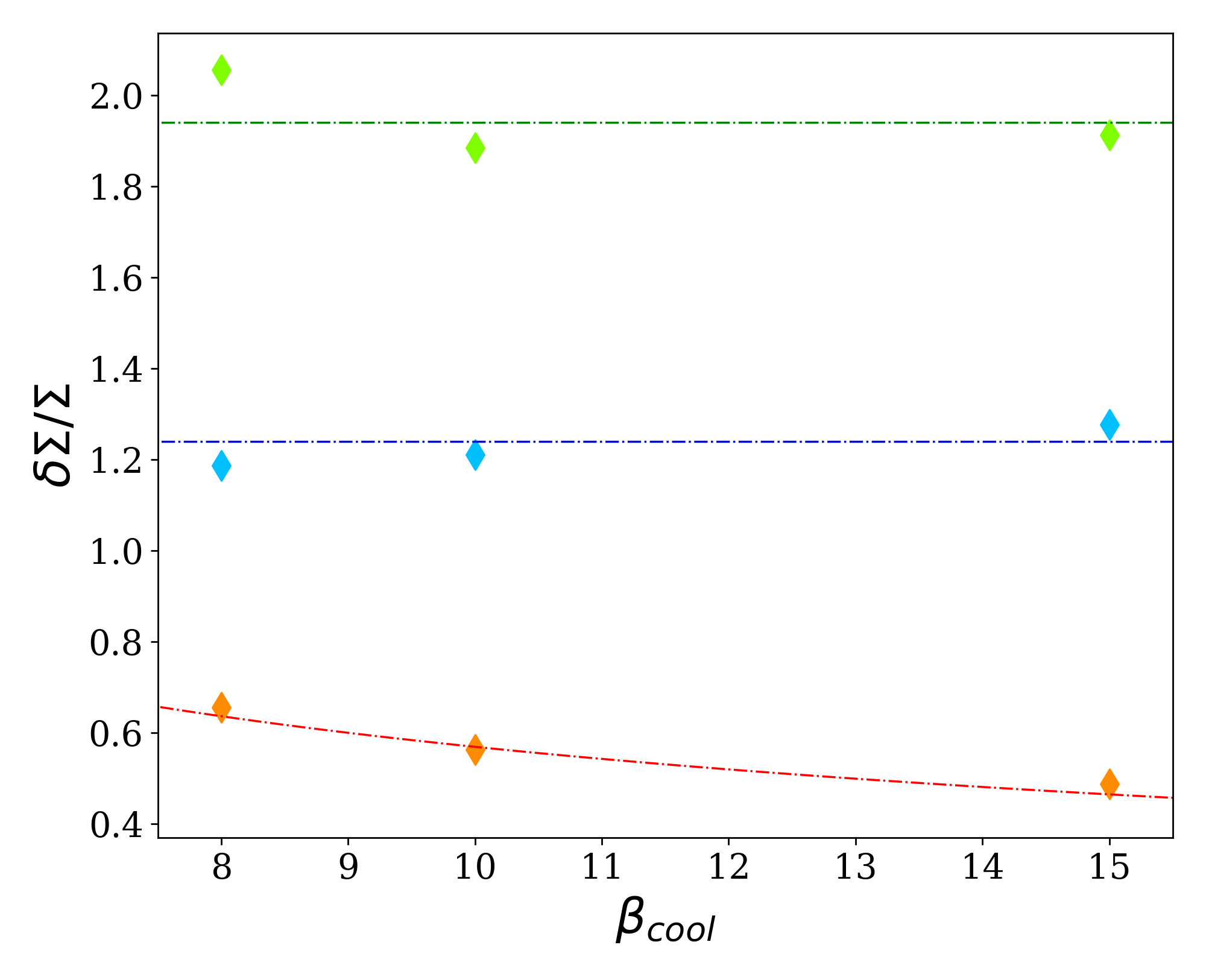}
    \caption{Density contrast $\delta\Sigma/\Sigma_0$ of gas (orange), large dust grains (blue) and small dust grains (green) as a function of the cooling factor, for $M_d/M_\star = 0.05$.}
    \label{deltasigma}
\end{figure}

GI spiral arms trap dust particles \citep{dipierrotrap}, since they are both pressure maxima and gravitational potential minima. We use the simulations to quantify how this phenomenon depends on the model parameters. Figure~\ref{epsilon_histo} shows the distribution of the dust to gas ratio for different $\beta_\text{cool}$ and $M_d/M_\star$, for a set of simulations with large dust particles. The initial value of the dust to gas ratio is $10^{-2}$. In Figure~\ref{epsilon_histo} the higher tail of the distributions reaches values of $\gtrsim 10^{-1}$, implying that dust concentration by up to approximately an order of magnitude is happening. Figure~\ref{epsilon_largesmall} shows a comparison between the  {dust to gas} ratio distributions of large and small particles. As expected, dust concentration in spiral arms is stronger for smaller particles, and it can approach values of the order of unity, since their aerodynamical coupling with gas is stronger.  The strength of dust trapping is determined by both the aerodynamic coupling between gas and dust and the gravitational potential of gas spiral arms. The combined effect of gravitational and drag interaction is maximised when $\text{St}\simeq Q\simeq 1$ \citep{zhu1}, thus, in our simulations, smaller particles reach higher values of the dust to gas ratio. No particular correlations are found between $\epsilon$ and the disc to star mass ratio, while there is a slight dependence on the cooling factor. In order to understand this relationship, Figure~\ref{deltasigma} shows the quantity $\delta \Sigma / \Sigma_0$ for gas (orange dots) and dust (large particles - blue dots, small particles - green dots) as a function of the cooling factor $\beta_\text{cool}$. The quantity $\Sigma_0$ is the azimuthally averaged surface density at a fiducial radius of 10~au, and the quantity $\delta\Sigma$ is its standard deviation. For the gas, it is known that $\delta\Sigma_g / \Sigma_{g0}\propto\beta_\text{cool}^{-1/2}$ \citep{cossins}, and we recover this behaviour in our simulations. For the dust, the situation is different. We do not find any evident correlation between the density contrast and the cooling factor. So, if we assume that 
\begin{equation}\label{deltasigmasigma}
    \frac{\delta \Sigma_g}{\Sigma_{g0}} \propto \beta_\text{cool}^{-1/2},\quad \frac{\delta \Sigma_d}{\Sigma_{d0}} = \text{const},
\end{equation}
the ratio between these two quantities is
\begin{equation}
    \frac{\delta \Sigma_g}{\Sigma_{g0}}\frac{ \Sigma_{d0}}{\delta\Sigma_{d}} = 0.01\epsilon^{-1} \propto \beta_\text{cool}^{-1/2},
\end{equation}
where $\Sigma_{d0}/\Sigma_{g0} = 1/100$, meaning that $\epsilon \propto \beta_\text{cool}^{1/2}$. This is the  positive correlation we found before. Why do we not find any evident correlation between the dust density contrast and the cooling factor?  Physically, the dust experiences the effect of the gas  cooling through gravitational and drag forces. When $\text{St}\ll 1$, dust and gas particles are indistinguishable, and so $\delta\Sigma_g / \Sigma_{g0} = \delta\Sigma_d / \Sigma_{d0} \propto \beta_\text{cool}^{-1/2}$. For higher Stokes number, the drag force is weaker, and the dust is less influenced by the gas cooling. In this case, we expect the relationship between $\delta\Sigma_d / \Sigma_{d0}$ and $\beta_\text{cool}$ to be flatter than $\beta_\text{cool}^{-1/2}$: if this condition is respected, the correlation between $\epsilon$ and $\beta_\text{cool}$ will be positive. In general, $\delta\Sigma_d / \Sigma_{d0}$ is a function of both the cooling factor and the Stokes number.

\subsection{Dust excitation: the $\xi$ parameter}

\begin{figure}
	\includegraphics[scale=0.475]{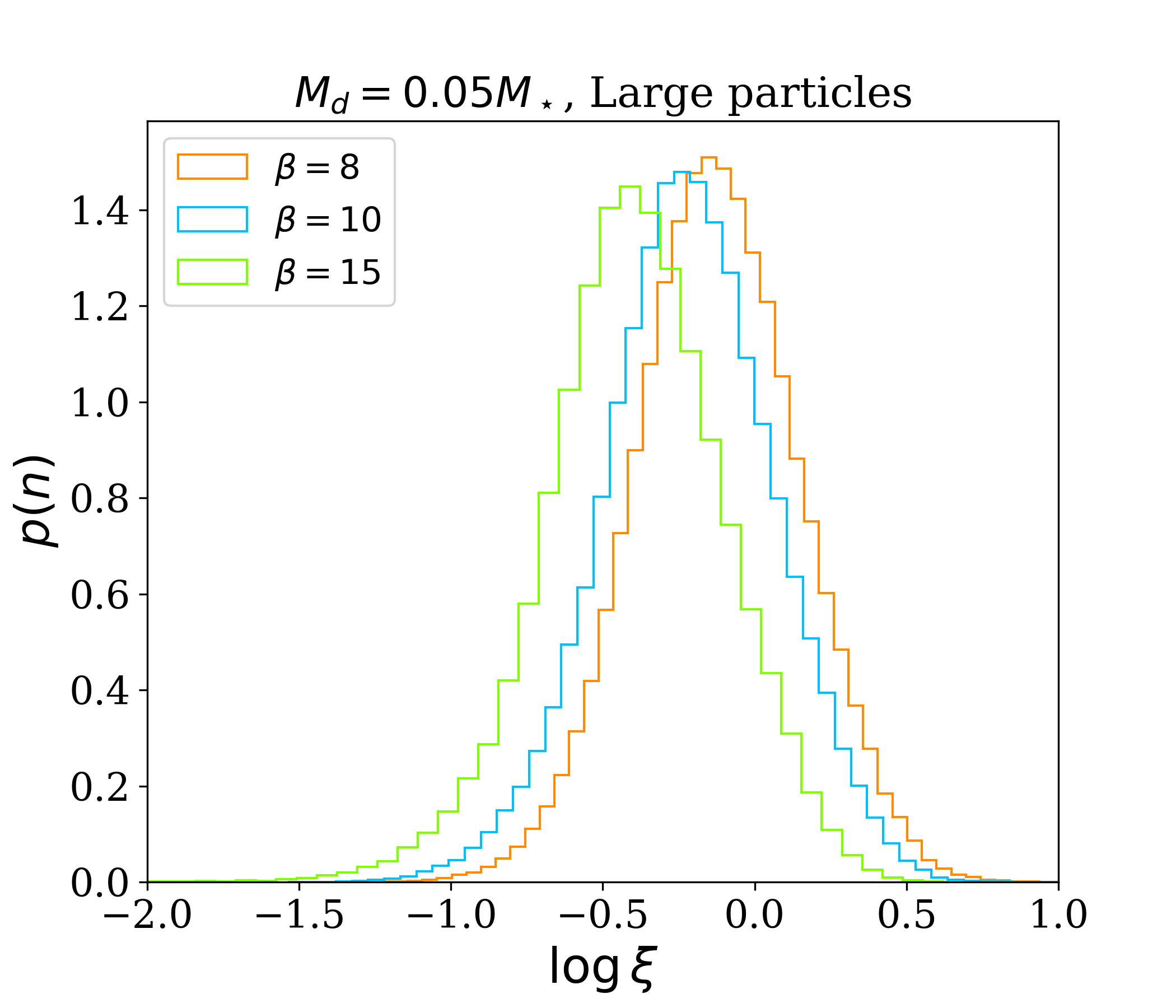}
	\includegraphics[scale=0.475]{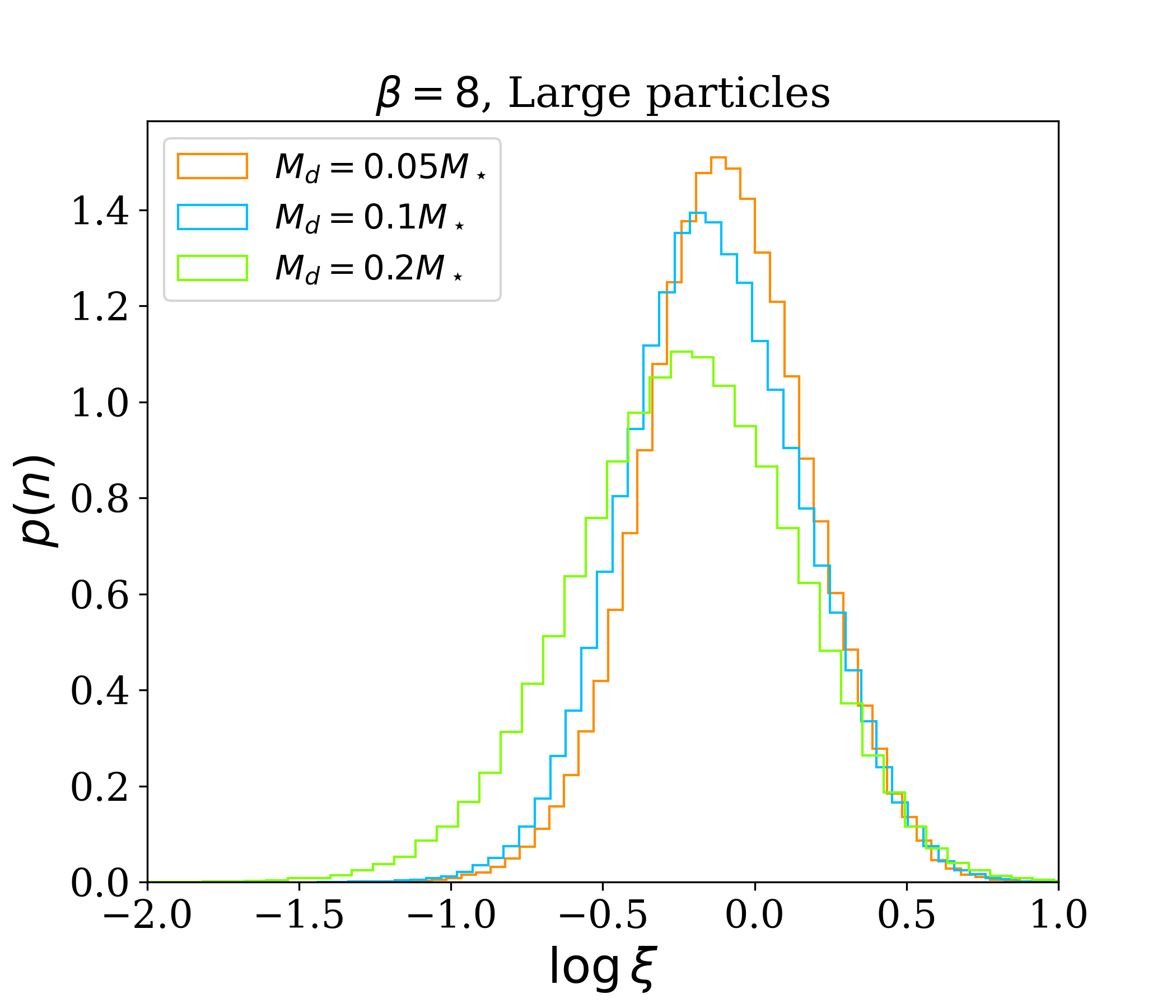}
    \caption{Distribution of the relative temperature for different values of cooling factor (top panel) and disc to star mass ratio (bottom panel). The simulations shown in these plots are \textbf{S1,S2,S3,S7,S13}.}
    \label{xi_histo}
\end{figure}

To investigate dust excitation by spiral arms, we study how the relative temperature $\xi = (c_d/c_g)^2$ varies as a function of the simulation parameters. Figure~\ref{xi_histo} shows the distribution of the dust relative temperature $\xi$ for different values of $\beta_\text{cool}$ and $M_d/M_\star$ for a set of simulations with large dust particles.  {We observe that for the simulations in which dust collapse is not happening, the dust dispersion velocity reaches very quickly ($<1$ outer orbital time) a steady value.} The relative temperature shows a negative correlation with both the disc-to-star mass ratio and the cooling factor. For the disc-to-star mass ratio, this can be understood by considering the relationship with the spiral morphology: $M_d/M_\star$ is inversely proportional to the azimuthal wavenumber $m$ \citep{cossins}, hence massive discs have fewer spiral arms. Since dust is excited because of ``spiral kicks'' \citep{kicks}, the lower the number of spiral arms, the less the dust is excited. For the cooling factor, the negative correlation can be understood in two ways. First, the cooling rate $\beta_\text{cool}$ is linked to the amplitude of the spiral perturbation according to  {eq.~\ref{deltasigmasigma}}. Since gas spiral arms excite dust particles by kicking them every passage, the higher is the perturbation, the stronger is the kick, and so the excitation. Second, in gravito-turbulent regime, transport of angular momentum is driven by the spiral perturbation, and it is possible to define an $\alpha-$viscosity coefficient related to the cooling rate  {(eq.~\ref{alphacool})}. The height of a dust layer is determined by the interaction with the gas and by the vertical diffusion. In the hypothesis that the vertical diffusion coefficient is equal to the azimuthal one, we can obtain the height of the dust layer as
\begin{equation}
    H_d = H_g \sqrt{\frac{\alpha}{\alpha+\text{St}}}.
\end{equation}
If we assume that $\alpha=\alpha_\text{GI}$, the dust layer height, and thus the dust dispersion velocity, is inversely proportional to $\beta_\text{cool}$.

\begin{figure}
	\includegraphics[scale=0.475]{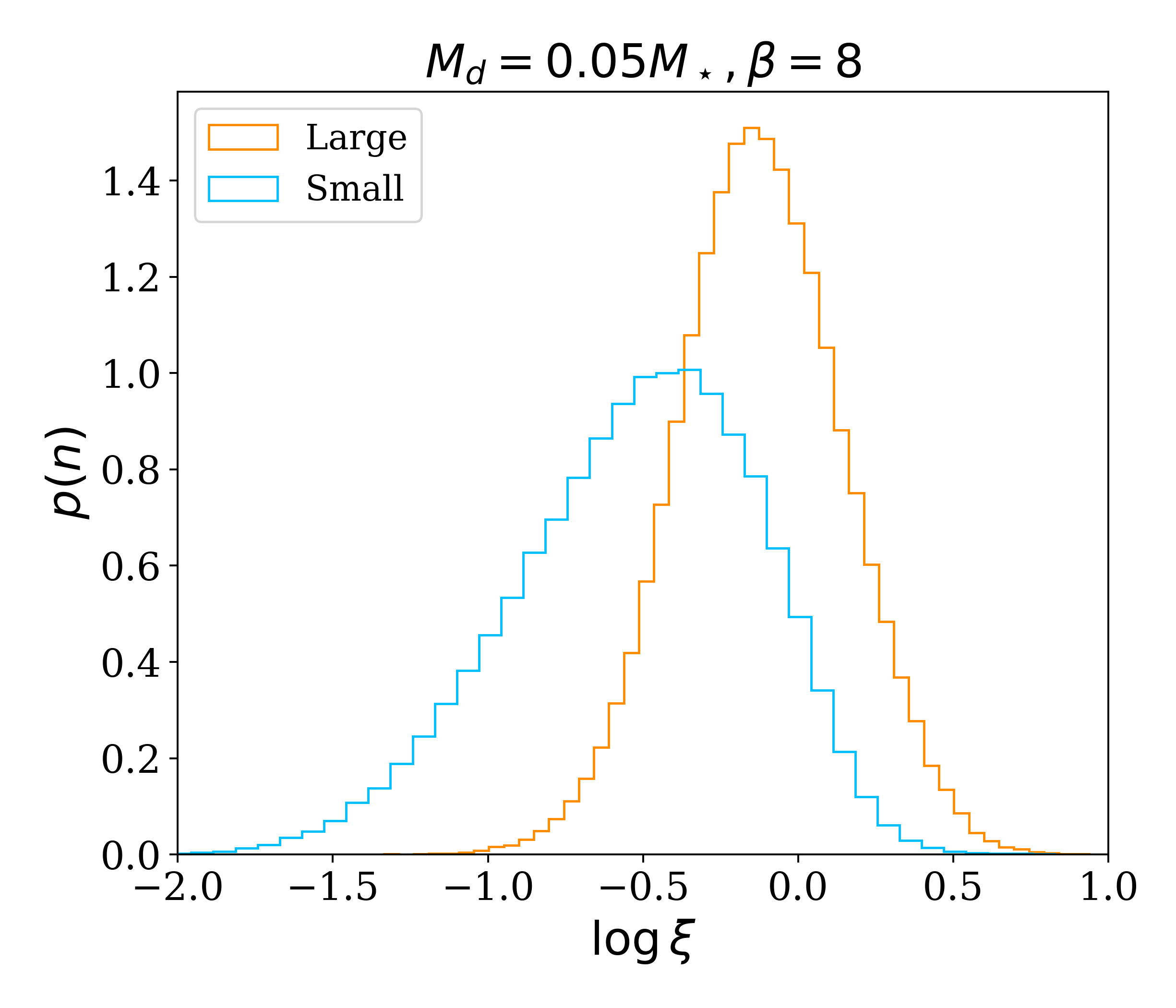}
    \caption{Comparison of the distribution of relative temperature of large (orange line) and small (blue line) dust particles. The simulations shown in this plot are \textbf{S1} and \textbf{S4}.}
    \label{comparison_xi}
\end{figure}

Figure~\ref{comparison_xi} compares the relative temperature distributions of large and small dust particles. Small particles are colder than the large ones, and their relative temperature is almost completely enclosed in the interval $\xi_\text{small}\in[10^{-2},1]$, meaning that their random motions are subsonic. On average, the  distribution of $\xi_\text{small}$ is shifted by one order of magnitude compared to $\xi_\text{large}$. This behaviour is in agreement with what \cite{boothclarke16} found: larger particles tend to be dynamically hotter, because the kicks of the gas spiral are more effective, while if the coupling with the gas is stronger, the kicks are damped because of the drag force. 

In principle, the observed trend would imply that increasing $\beta_\text{cool}$ would lead to an arbitrarily thin dust layer, eventually causing gravitational collapse. {This would be a direct analogue of the classical \cite{Goldreich73} mechanism for planetesimal formation in a (weakly) self-gravitating context, and in a more realistic model it likely be limited in a similar way by the excitation of shear turbulence \citep{cuzzi93}}. However, there is an upper limit for $\beta_\text{cool}$ set by the value above which the transfer of angular momentum would be driven by some process other than gravitational instability. To compute an estimate of the maximum cooling time, we require $\alpha_\text{GI}$ to be larger than $10^{-3}$. Observations of protoplanetary discs that are not expected to be self-gravitating suggest that this is a reasonable upper limit to the strength of turbulence \citep{flaherty}. Thus, we obtain
\begin{equation}
    \beta_\text{cool}^\text{max} = 400\left(\frac{\alpha}{10^{-3}}\right)^{-1},
\end{equation}
that corresponds to a minimum density perturbation
\begin{equation}
    \left.\frac{\delta \Sigma_g}{\Sigma_{g0}}\right|_\text{min} = 0.05\left(\frac{\alpha}{10^{-3}}\right)^{1/2}.
\end{equation}

\begin{table}
\caption{Table that summarizes the correlations between $\epsilon,$ $\xi$ and $\beta_\text{cool},$ $M_d/M_\star$ and St.}
\centering
\begin{tabular}{|l|l|l|l|}
\hline
\multicolumn{1}{|r|}{\textbf{}} & \textbf{$\beta_\text{cool}$} & \textbf{$M_d/M_\star$} & \textbf{St} \\ \hline 
\textbf{$\epsilon$} & Positive & None & Negative \\ \hline
\textbf{$\xi$} & Negative & Negative & Positive \\ \hline
\end{tabular}\label{table1}
\end{table}

Table~\ref{table1} summarizes the relationships we have discussed in these paragraphs. A broader collection of histograms can be found in Appendix~\ref{appb}.

\begin{figure*}
	\includegraphics[scale=0.435]{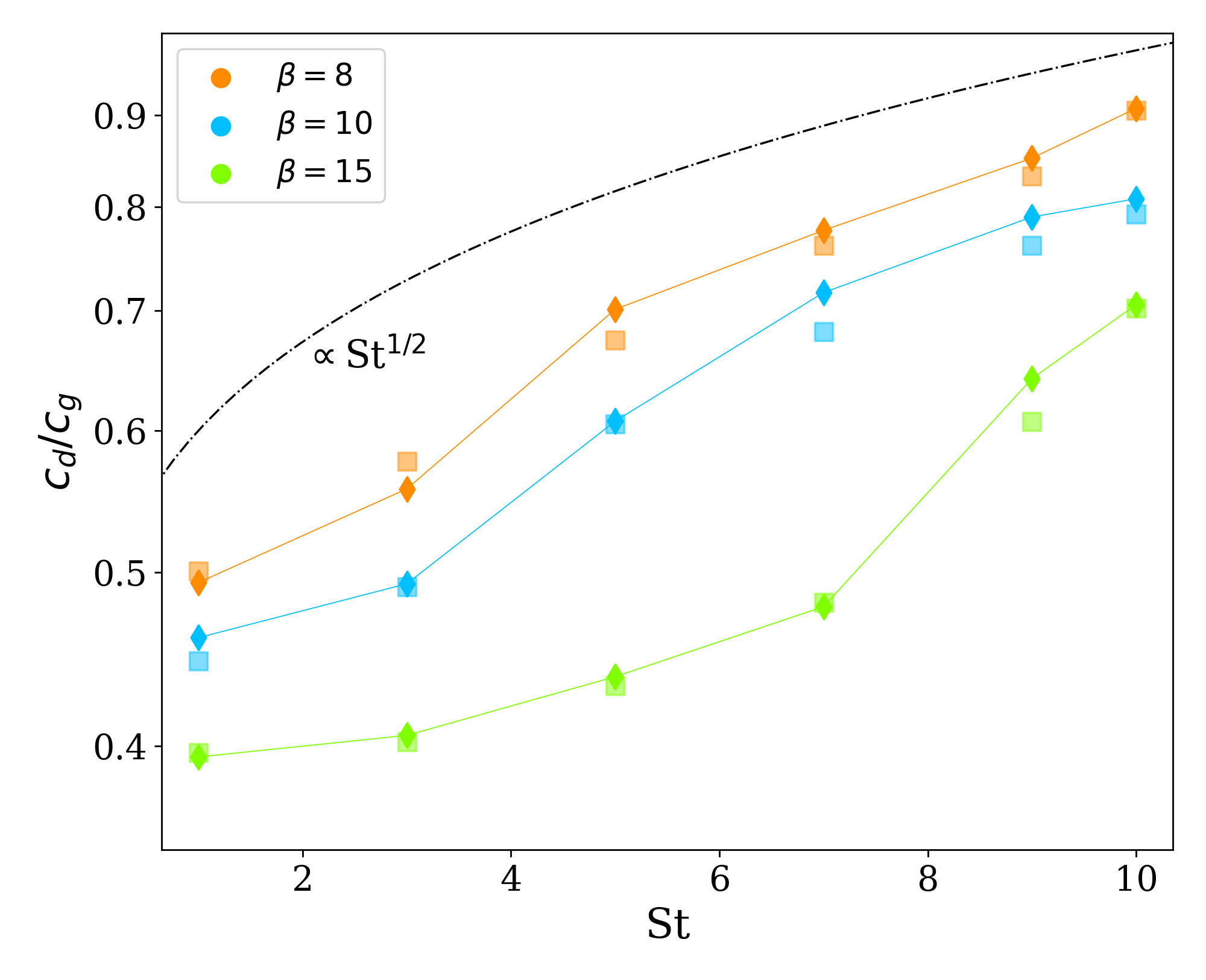}
	\includegraphics[scale=0.435]{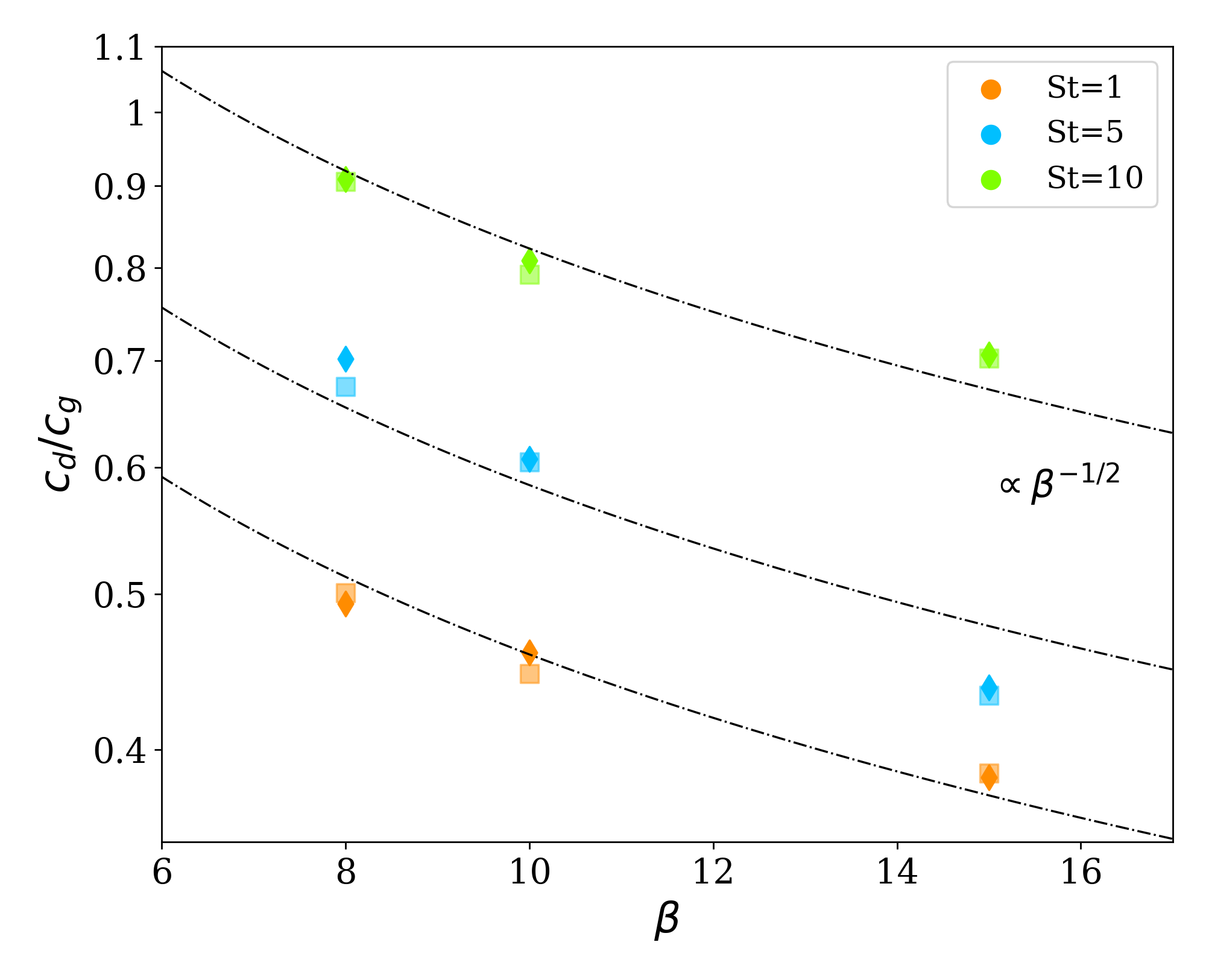}
    \caption{Comparison with \citet{boothclarke16}. The left panel shows how dust dispersion velocity depends on the Stokes number, for different values of cooling factor, compared to the expected relationship $\propto \text{St}^{1/2}$. The right panel, shows how dust dispersion velocity depends on the cooling factor, for different values of Stokes number, compared to the expected relationship $\propto \beta_\text{cool}^{-1/2}$. All the simulations shown in these plots have $M_d/M_\star=0.05$. The diamonds are the values obtained from the standard resolution simulations, while the squares from the high resolution ones.}
    \label{xi_cose}
\end{figure*}

\cite{boothclarke16} studied the relationship between the dust excitation and both the cooling and the Stokes number. They found that $c_d\propto\beta^{-1/2}\text{St}^{1/2}v_k$, where $v_k$ is the Keplerian speed. To compare with \cite{boothclarke16}, we use our data to reproduce Figures \textbf{7} and \textbf{13} of their paper, that show a relationship between the dust velocity dispersion and $\beta_\text{cool}$ and St. To do so, we divided the particles into equally spaced intervals of Stokes number and, for each particle, we computed the mean value of $c_d/c_g = \sqrt{\xi}$. The comparison is shown in Figure~\ref{xi_cose}, where we show the results of simulations with $M_d/M_\star=0.05$, for both standard and high resolution cases. The previously derived relationships with the Stokes number (the left panel) and with the cooling factor (the right panel) are well recovered. Using our two fluid algorithm, it is too computationally expensive to analyse properly the case of $\text{St}<1$. However, in this case we expect that as the aerodynamical coupling with the gas is stronger, $c_d/c_g$ should increase, eventually reaching $c_d=c_g$ for $\text{St}\to0$. This growth for $\text{St}<1$ has been shown by \cite{boothclarke16}.

\subsection{Two fluid instability}
\begin{figure*}
	\includegraphics[scale=0.45]{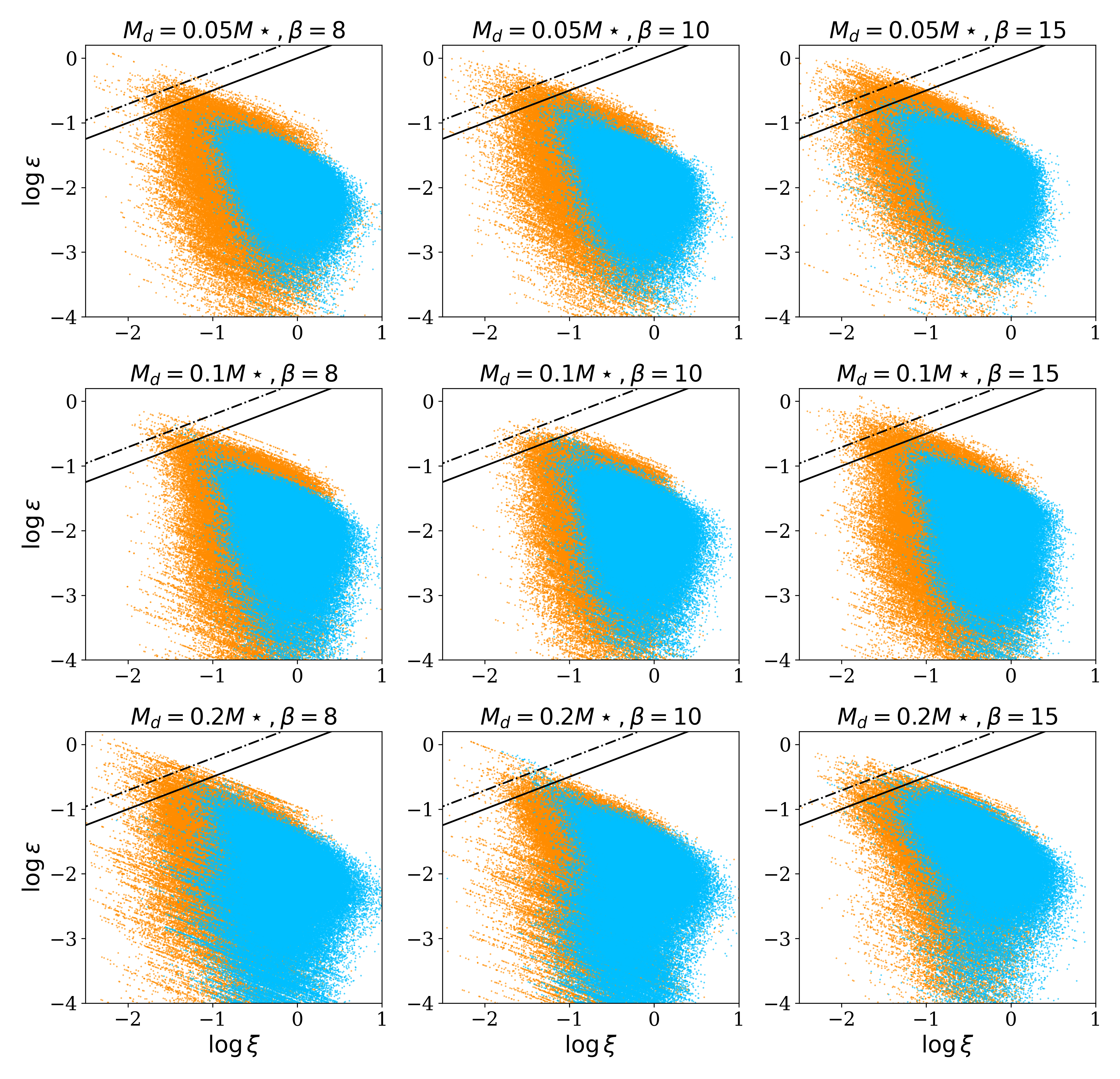}
    \caption{Distribution of large (blue) and small (orange) dust particles in the $(\epsilon,\xi)$ diagram, for different values of disc-to-star mass ratio and cooling factor.  {The black lines corresponds to the dust driven GI threshold for $\text{St}=\infty$ (eq.~\ref{dustdr}, solid line) and for $\text{St}=0.5$ (eq.~\ref{dustdrdrag}, dashed line). We choose $\text{St}=0.5$ as a minimum value since the number of particles with Stokes number lower than this is negligible. The particles above the region are in a dust driven GI regime.}}
    \label{instabilitydistrib}
\end{figure*}

\begin{figure*}
	\includegraphics[scale=0.4]{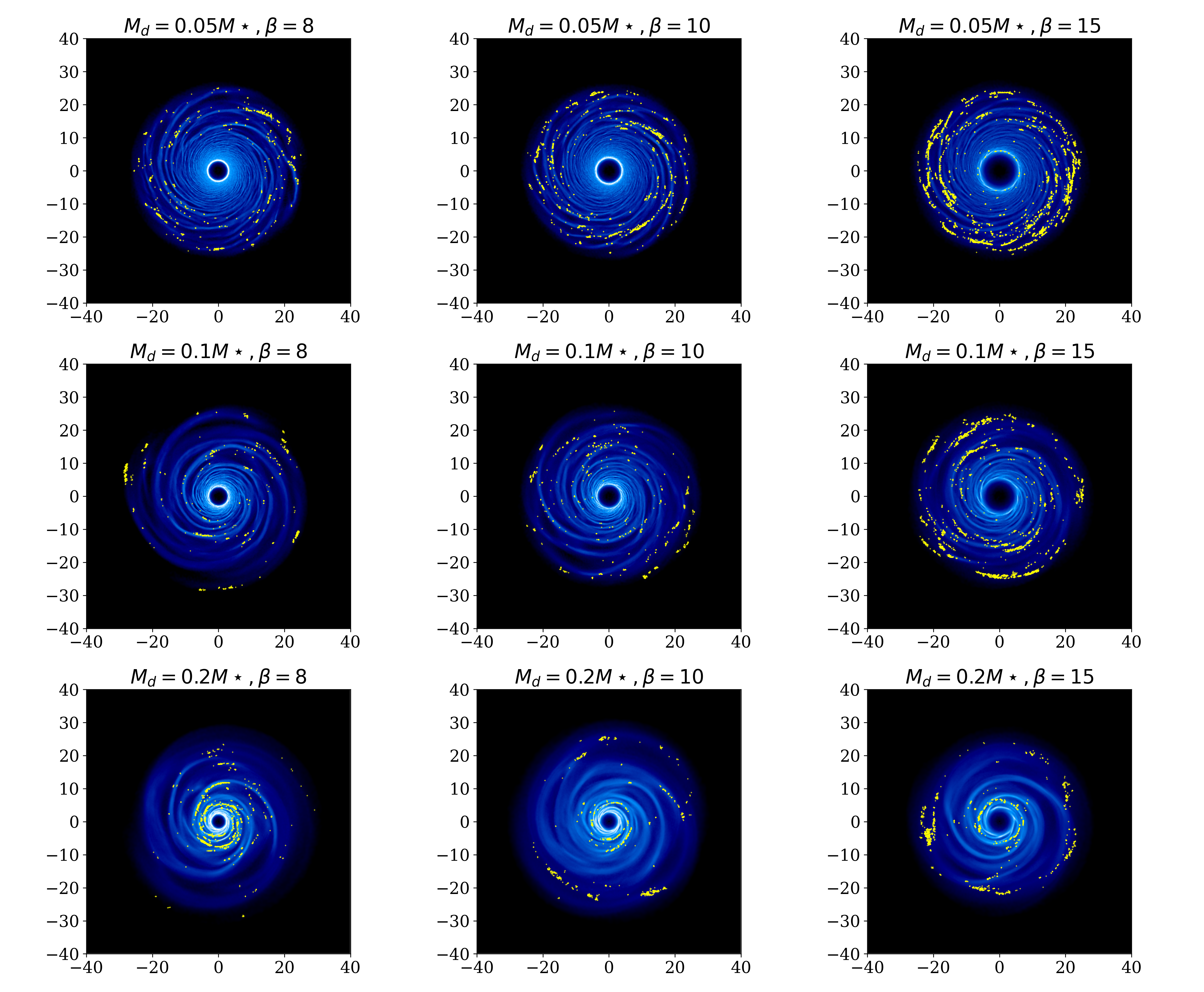}
    \caption{Total density maps for small dust particles simulations (S4,S5,S6,S10,S11,S12,S16,S17,S18 in order). Yellow dots correspond to particles for which the instability is dust driven.}
    \label{instapart}
\end{figure*}

In this section, we apply the two fluid instability theory presented in Section~\ref{s2}. The gas-only and gas-and-dust  models for gravitational instability have both been developed within a linear framework; hence, in principle, the quantities $\epsilon$, $\xi$ and St should be evaluated in the unperturbed state. However, in this work, we are evaluating them in the perturbed one. Although not completely self-consistent, it gives us an idea of the most unstable regions of the disc. Figure~\ref{instabilitydistrib} shows the distribution of large (blue) and small (orange) dust particles in the $(\xi, \epsilon)$ diagram:  {the black lines corresponds to the dust driven GI threshold for $\text{St}=\infty$ (eq.~\ref{dustdr}, solid line) and for $\text{St}=0.5$ (eq.~\ref{dustdrdrag}, dashed line). We choose $\text{St}=0.5$ as a minimum value since the number of particles with Stokes number lower than this is negligible. The particles above the region are in a dust driven GI regime.} We find that the number of small particles for which the instability is dust driven is greater compared to large ones: this is because small particles have both larger dust to gas ratio and lower dispersion velocity. In addition, the number of dust driven particles increases with the cooling factor and with the disc to star mass ratio, as already discussed in previous sections. To understand the spatial location of these particles in the disc, Figure~\ref{instapart} shows the particles that satisfy condition \ref{dustdr} superimposed on the total density map. As expected, the most unstable regions of the disc are not randomly distributed, but correspond with the spiral arms. 

\begin{figure}
	\includegraphics[scale=0.41]{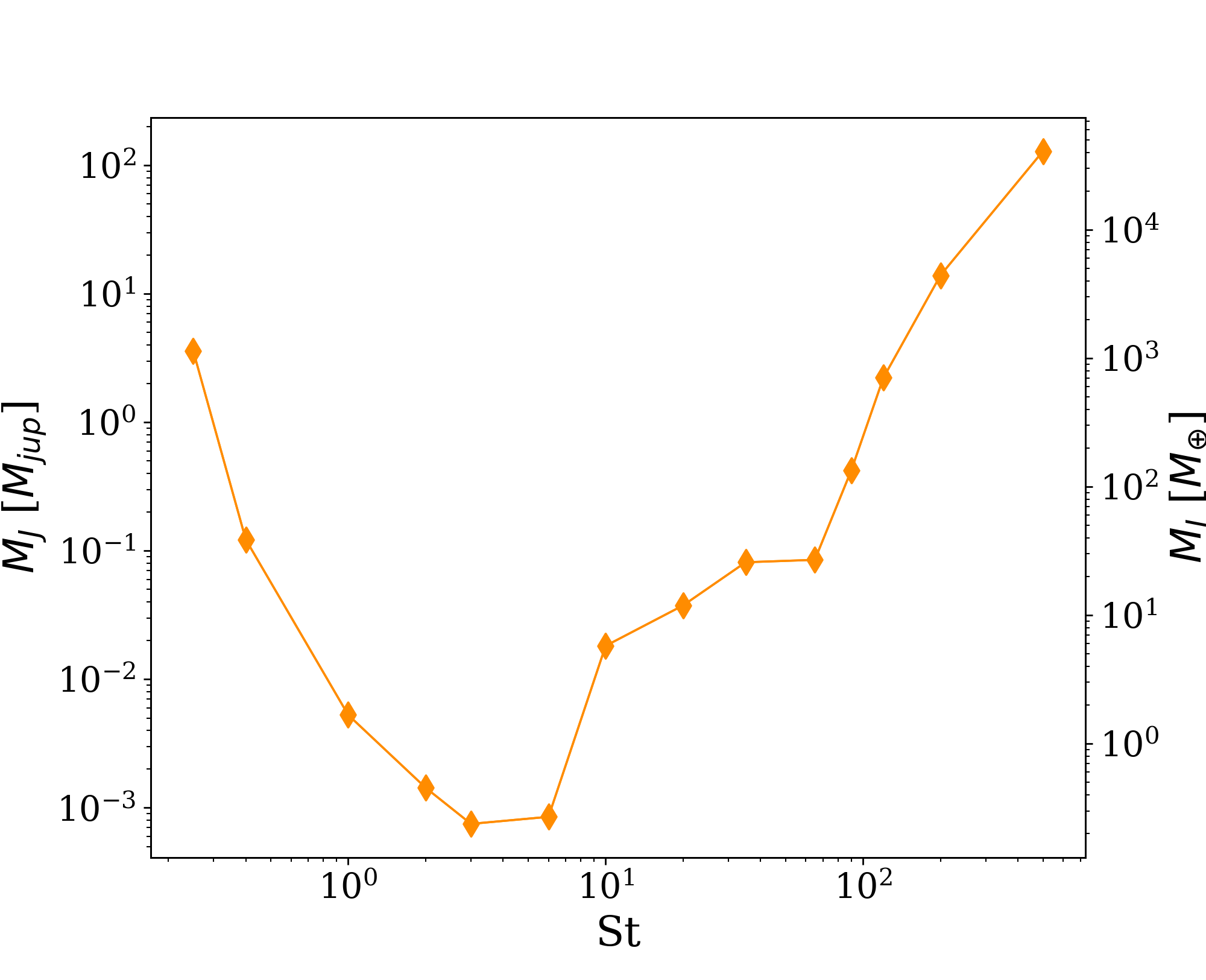}
    \caption{Jeans mass given by eq. (\ref{jmass_complete}) evaluated at each dust particle location as a function of the Stokes number, for $M_d/M_\star =0.05$ and $\beta_\text{cool} = 8$.}
    \label{jeans_stokes}
\end{figure}

Figure~\ref{jeans_stokes} shows the value of the Jeans mass for $M_d/M_\star=0.05$ as a function of the Stokes number. The curve can be divided into two parts: for $\text{St}\in(0,5)$, the Jeans mass is decreasing with the Stokes number, reaching its minimum at about $\text{St}\sim 3$. This happens because for $\text{St}\to 0$, the particles are indistinguishable, and the instability is gas driven. It is important to notice that the number of particles with small Stokes number is low, hence the binning in Stokes number presents a considerable scatter. By increasing the Stokes number, $\xi$ decreases and the two fluids behave more and more differently. When the Stokes number is approximately 1, the relative temperature is a minimum and the dust to gas ratio is high, so the instability becomes dust driven. Otherwise, for $\text{St}> 5$, the Jeans mass increases with the Stokes number. Indeed, the relative temperature increases, and the dust to gas ratio decreases. Hence, the system transitions from dust into gas driven instability, again, eventually reaching the gas-only component model value.

\section{Discussion}\label{s5}

\subsection{Dust collapse}

\begin{figure}
	\includegraphics[scale=0.415]{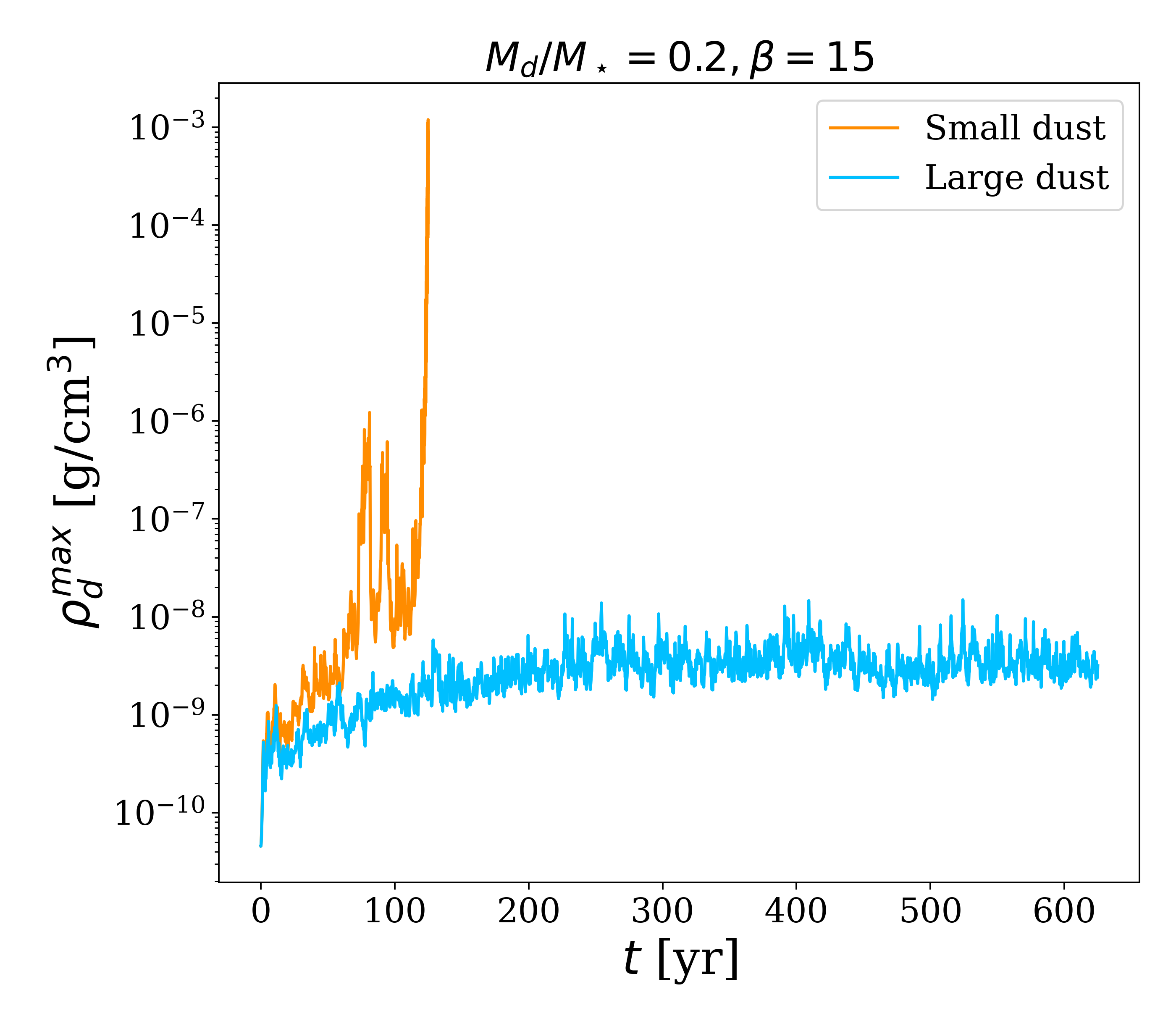}
	\includegraphics[scale=0.415]{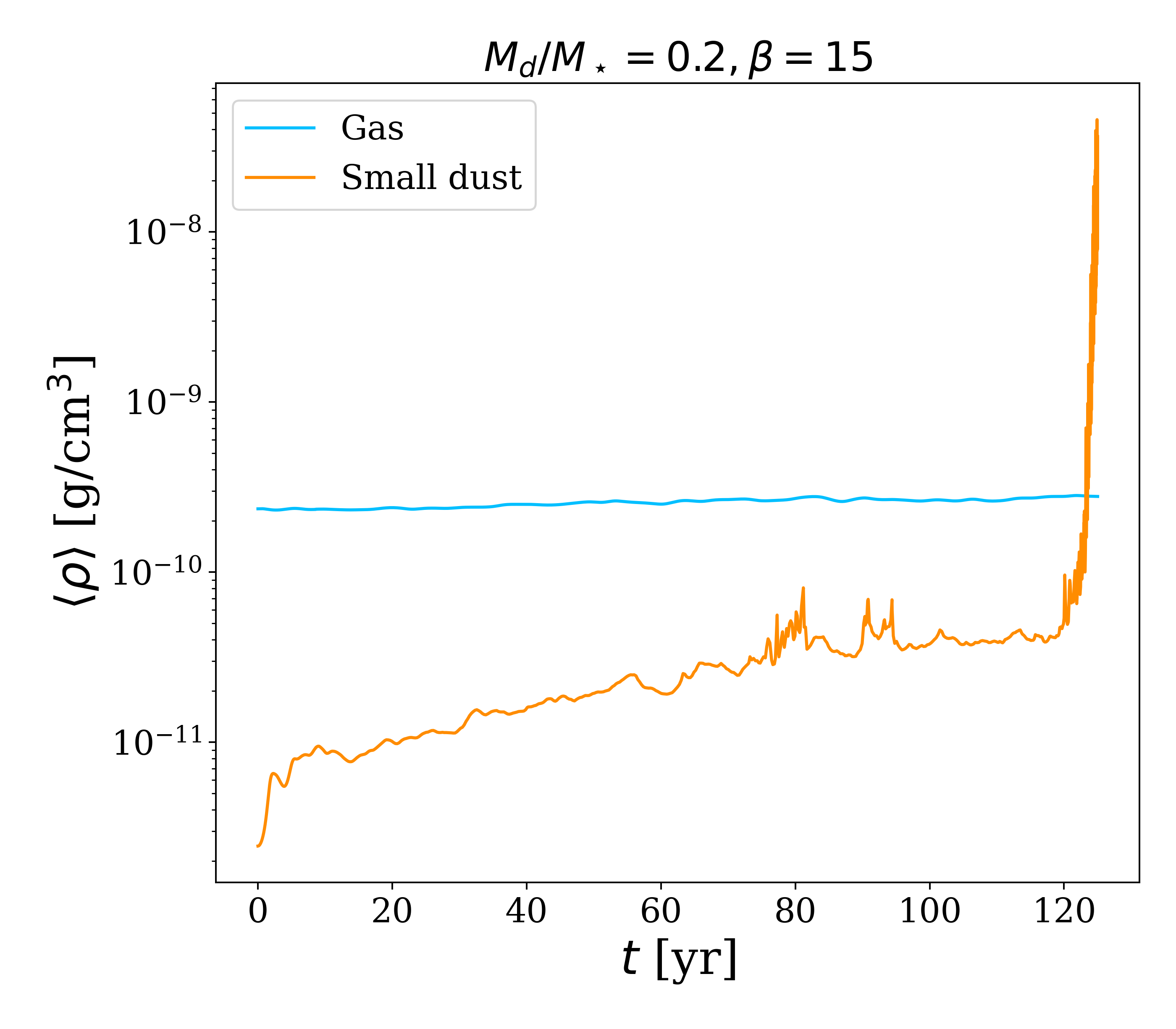}
    \caption{Maximum dust density as a function of time for simulations S15 ($M_d/M_\star = 0.2, \beta=15$, large dust particles, blue line) and S18 ($M_d/M_\star = 0.2, \beta=15$, small dust particles, blue line).}
    \label{max_density}
\end{figure}

The simulations with $M_d/M_\star=0.2$ and small dust particles do not reach 5 outer orbits, since the simulation stops due to the onset of dust collapse. This happens because the stopping time of collapsing dust particles becomes smaller than the time step of the code. Indeed, the stopping time is inversely proportional to the total density $\rho_\text{tot} = \rho_g + \rho_d$, and since the dust density is increasing, because of the collapse, the stopping time tends to zero. The top panel of Figure~\ref{max_density} shows the maximum dust density as a function of time for small and large dust particles, in the run with $M_d/M_\star = 0.2$ and $\beta=15$. While large dust particles reach a quasi-steady state, the small particle density exponentially increases in the first orbit. This is the signature of dust collapse. This phenomenon is visible in simulations \textbf{S16}, \textbf{S17} and \textbf{S18}, and happens only in the dust component. The bottom panel of Figure~\ref{max_density} shows a comparison between gas and small dust averaged density as a function of time, for $M_d/M_\star = 0.2$ and $\beta=15$. At $t=0$, $\langle\rho_d\rangle=10^{-2}\langle\rho_g\rangle$. Whereas the gas average density is constant with time, the dust density increases because of dust trapping, eventually exceeding that of the gas. This means that any clumps forming from this mechanism would be substantially made up of solids, and would likely be identified with the rocky core of a giant planet.  {However, in this work we do not want to characterize the outcome of this collapse, which is a complex topic. Indeed, simulations of planetesimal collapse }\citep{nesv}  {show that a rotating self-gravitating cloud of dust does not monolithically collapse, meaning that it is not possible to directly equate the cloud mass with the planetary core one.}

To identify and analyse dust clumps in more detail, we define the numerical conditions that should be respected for a clump to be physical and not affected by resolution. For a clump radius $r_\text{clump}$ the smoothing length of the dust particle $h_i$ should be less than a fraction of the clump radius, in order to be resolved. This condition translates into $h_i < \eta r_\text{clump}$, where $\eta$ is less than unity. We take $\eta = 1/2$. Physically, a gravitationally bound clump is a collection of particles whose thermal support does not balance the gravitational one. We define the thermal and gravitational energy of particles inside $r_\text{clump}$ as follows:
\begin{equation}
    e_\text{th} = \sum_i^{i\in r_\text{clump}} m_i c_i^2,
\end{equation}
\begin{equation}
    e_\text{gr} = -\frac{1}{2}\sum_{i}^{i \in r_\text{clump}}\sum_{j=i}^{j \in r_\text{clump}}m_im_j[\varphi(r_{ij},h_i) + \varphi(r_{ij},h_j)],
\end{equation}
where $\varphi$ is the gravitational softening kernel and $r_{ij} = |\mathbf{r}_i-\mathbf{r}_j|$. We used a cubic spline softening kernel, and the detailed expression can be found in the appendix of \cite{softening}. According to the virial theorem, if the force between any two particles can be described in terms of  a potential energy $\Phi\propto r^{n}$, where $r$ is the distance between two particles, the equilibrium state respects the following condition
\begin{equation}
    2\langle T\rangle = n \langle \Phi \rangle,
\end{equation}
where $T$ is the kinetic energy. In the case of gravitational interaction, the virial theorem reads $\langle T\rangle / \langle \Phi \rangle = -1/2$. We define a clump as a region of the space where the dimensionless quantity $\alpha_J=-e_\text{th}/e_\text{gr} < 1/2$. Then, in order to be sure that the collapse is physical, and not artificial, we verify that $h_g<h_d$ in the region where there is the dust clump. The last condition requires that the resolution of the gas in the region where there is a possible clump should be higher compared to the dust resolution.

To summarize, the conditions under which we define a dust clump are the following
\begin{itemize}
    \item $h_{g,i} \leq h_{d,i}$,
    \item $h_{d,i} < \eta r_\text{clump}$,
    \item $\alpha_J < 1/2$.
\end{itemize}

\begin{table}
\caption{Comparison between the expected and the observed Jeans Mass in the simulations where dust collapse happens.}
\centering
\begin{tabular}{|l|l|l|}
\hline
\multicolumn{1}{|r|}{\#} & \textbf{Expected $M_J$ [M$_\oplus$]} & \textbf{Simulation [M$_\oplus$]}  \\ \hline 
\textbf{S16} & 2  & 0.6   \\ \hline
\textbf{S17} & 2.2 & 1.  \\ \hline
\textbf{S18} & 4.5 & 3.2  \\ \hline
\end{tabular}\label{t2}
\end{table}

In simulations \textbf{S16}, \textbf{S17} and \textbf{S18} there are particles that respect the previous conditions, implying that dust collapse has happened. Table~\ref{t2} shows the mass of the clumps, obtained by summing the mass of each particle that is gravitationally bound and the one obtained from analytical theory from Eq.~\ref{jmass_complete}. The masses are broadly in agreement with analytic expectations. For \textbf{S16}, for example, the derived Jeans mass is of the order of an Earth mass. In general, the mass of the clump computed from the simulation is smaller compared to the one expected from the analytical theory: this is not surprising, since with the simulation we are only able to appreciate the initial phase of the collapse. Indeed, as soon as the stopping time is smaller than the time step, the simulation stops. To avoid this problem, one could decrease the time step of the code, but it is computationally expensive. Otherwise, one could not consider the dust density when computing the stopping time: this procedure has been applied in previous works to increase the velocity of the simulations \citep{poblete19,longarini21} but this approximation is valid only for  {small dust to gas ratios}. While there this was justified, here this is not possible, since dust is collapsing and the dust to gas ratio becomes higher than unity. Finally, to study the early evolution of clumps with an SPH code, it would be possible to simulate them as sink particles: so far, the creation of dust sink particles is not possible in \textsc{phantom}.


\subsubsection{Gas-dust coupling during the collapse}
The bottom panel of Figure~\ref{max_density} shows that the collapse happens only in the dust component, and the gas is not influenced. This could sound surprising: indeed, in high density regions, we expect the stopping time (eq.~\ref{tstop}) to be small.
So, why is the dust collapse not influencing the gas? The degree of coupling is measured with the Stokes number, that compares the strength of the drag force with the ones that are acting on the particle. Usually, in a flat protoplanetary disc, the Stokes number is computed as the ratio of the stopping time and the dynamical time, that is the typical timescale of a particle orbiting around a central object at a distance $R$. However, in this situation, the dust clump is driving the dynamics of the surrounding particles, and hence we should compare the stopping time with the free fall time in order to understand the degree of coupling of particles. The typical timescale of the infall of a spherically-symmetric distribution of mass is
\begin{equation}
    t_\text{ff} = \sqrt{\frac{3\pi}{32G\rho}}.
\end{equation}

Comparing the stopping time and the free fall time in the simulations where dust collapse is happening (Table~\ref{times}), we obtain that the particles in the collapsing region are uncoupled, since the ratio between the two timescales is higher than one.

\begin{table}{%
\caption{Stopping, dynamical and free fall timescales for the simulations that show dust collapse}\label{times}
\begin{tabular}{llllll}
\# & $t_s$ [yr] & $t_\text{dyn}$ [yr] & $t_\text{ff}$ [yr] & $t_s/t_\text{dyn}$ & $t_s/t_\text{ff}$ \\ \hline
\textbf{S16} & 2.7 & 7.9 & 0.4 & 0.3 & 6 \\
\textbf{S17} & 5.6 & 14.1 & 0.7 & 0.4 & 8 \\
\textbf{S18} & 7.2 & 17.7 & 1.6 & 0.4 & 4
\end{tabular}%
}
\end{table}

It is possible to quantify the critical density a clump should reach so that $t_\text{ff} < t_\text{dyn}$, that is
\begin{equation}
    \rho_\text{crit} = \frac{3\pi}{32}\frac{M_\star}{R^3}:
\end{equation}
when a clump reaches this density, the evolution of the surrounding particles is determined by the clump, and not by the star anymore. From that point, the aerodynamical coupling should be quantified by taking the ratio between the stopping time and the free fall time. Hence, for $\rho<\rho_\text{crit}$, $\text{St}\propto\rho^{-1}$, since the dynamical time does not depend on the density. For $\rho > \rho_\text{crit}$, the scaling changes since the free fall time depends on the density $\text{St}\propto\rho^{-1/2}$, and so does the degree of coupling.

\subsection{Gravitational instability in the context of protoplanetary disc evolution}
In this work, we focused on cooling driven  {GI}, but it is also possible to trigger it through infall. \cite{kratt1} found that in the infall-driven case the strength of the spiral perturbation is controlled by two dimensionless parameters, a thermal one, that relates the infall mass accretion rate $\dot{M}_\text{inf}$ to the characteristic sound speed of the disc, and a rotational one, that  compares the relative strength of rotation and gravity in the core. Obviously, the higher the accretion rate, the stronger is the spiral perturbation in the disc. We can naively associate the dust evolution we have modelled in the limit of fast (slow) cooling with high (low) infall rate. This association is expected to be  qualitatively correct, however a detailed comparison between these two regimes is given by \cite{krattlod}.

\begin{figure*}
	\includegraphics[scale=0.26]{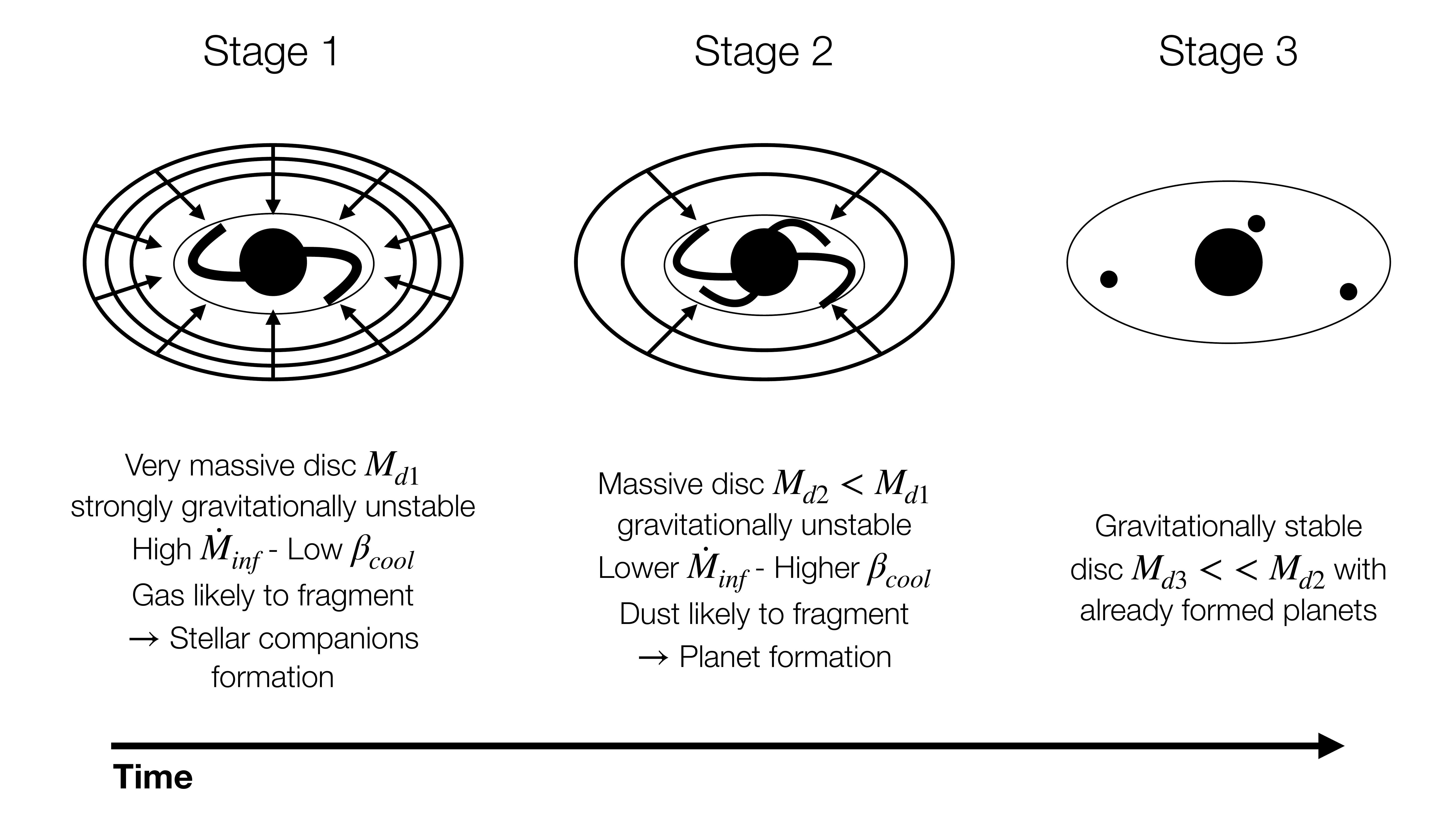}
    \caption{Scheme of the three stages of disc lifetime}
    \label{gas_th}
\end{figure*}

By studying the non-linear evolution of gas and dust in protoplanetary discs, we found that the instability conditions for the two components are different. It is well established that spiral fragmentation occurs in fast cooling gas discs, and it is possible to define a critical value of $\beta_\text{cool}$ below which fragmentation occurs. Simulations of cooling-driven fragmentation  \citep{Gammiecool,fragrice,lodclarke,merubate12} currently suggest that $\beta_\text{min}\simeq 3$ \citep{dengfrag}. For the dust, \cite{boothclarke16} found that dust becomes more unstable for higher $\beta_\text{cool}$, and we confirm this trend with 3D simulations. The differing behaviour of gas and dust suggests an interesting evolution in the outcome of gravitational instability within protoplanetary discs. During a first stage, at the beginning of the disc lifetime, we expect a very massive disc system characterized by strong  {GI}, caused by the high infall rate from the molecular cloud. If conditions allow gas fragmentation, because of the high Jeans mass, low mass stellar companions can be formed \citep{kratt1}. A second stage is characterized by a less massive disc, with lower infall rate, or equivalently longer cooling time. If conditions during this epoch trigger gravitational instability, it will lead to {\em dust-driven} fragmentation that could be responsible for the formation of rocky cores of giant planets. Then, in the third stage, there is a protostar surrounded by a planet hosting disc, characterized by substructures such as gaps, rings or planetary spirals. In this stage,  {GI} is not effective anymore because the disc mass is small and the transport of angular momentum is controlled by disc winds \citep{Tabone22} or other, non-self-gravitating, sources of turbulence \citep{Lesur22}.  {A schematic view of these stages is given by figure~\ref{gas_th}.}


\subsection{Application to an actual case: HL Tau}
HL Tau is a young ($< 1\text{Myr}$) protostellar system that shows axisymmetric structures (gaps and rings) in dust continuum emission \citep{HLTAU_1st}. The origin of rings and gaps is usually attributed to planet disc interaction \citep{linpapa}, with \cite{HLtaudipierro} finding that three protoplanets with masses $M_{p1} = \SI{61}{M_\oplus}$, $M_{p2} = \SI{83}{M_\oplus}$, $M_{p3} = \SI{170}{M_\oplus}$ at $R_{p1}=13.2$au, $R_{p2}=32.3$au and $R_{p3}=68.8$au best match the observations.
The formation of super-Earth mass planets at large radii, in such a young system, is a challenge for core accretion theory. If planets can form via the dust-induced collapse mechanism we have discussed in this work, HL~Tau is a plausible example of what the resulting planetary system might look like. We note that the inferred trend of increasing mass with radius can be interpreted within our model. In a simplistic way, we assume that $\text{St}\propto \Sigma^{-1}\propto R$, and that $\xi \propto \beta_\text{cool}^{-1}\text{St}\propto R^{4.5}R=R^{5.5}$, where we have supposed a realistic cooling law, according to which $\beta_\text{cool}\propto R^{-4.5}$ \citep{rafikovcool,clarkecool}. Supposing that the dust to gas ratio is constant, the Jeans mass of the gas-and-dust fluid model will be an increasing function of the radius, since the relative temperature is smaller for particles closer to the central star. So, within this hypothesis, we expect that the mass of the dust driven GI protoplanets will be an increasing function of the radius.  {We want to point out that this is just a qualitative argument. Indeed, to thoroughly investigate whether HL Tau planets can be formed through dust driven GI, one should properly model the system. In addition, planet migration and planet accretion should also be considered.}

\section{Conclusions}\label{concl}
Self-gravitating gas discs may be ubiquitous during the Class 0/I phases of YSO evolution \citep[e.g.][]{Lin90,Xu2021}. Observations suggest that the self-gravity can in some systems be strong enough as to trigger fragmentation \citep{Tobin2016}, while in other cases, such as the massive disc of the more evolved IM Lup system \citep{lodato23} the instability is expected to be more gentle. If that fragmentation occurs in the gas, as in the L1448~IRS3B system, the outcome is typically star or brown dwarf formation. Lower mass objects can be formed if the fragmenting fluid is instead the solid component of a two-fluid self-gravitating disc, given disc conditions that allows the collisional growth of dust to small macroscopic dimensions while the disc remains self-gravitating. Analytic estimates suggest that planetary cores of $\sim 1 - 10 \ M_\oplus$ could form from this mechanism at large orbital radius, with properties that could be identified with the population of ALMA-inferred disc-embedded planets \citep{Andrews2020}.

In this paper, we presented the results of SPH simulations of gas and dust in protoplanetary discs, studying the role of aerodynamic coupling in the context of gravitational instability. We analysed our results in the framework of two fluid gravitational instability, and compared our findings with previous numerical works, finding generally good agreement.  

Our main results can be summarized as follows:
\begin{enumerate}
    \item We studied the relationship between the dust to gas ratio $\epsilon$, the relative temperature between gas and dust $\xi$ and the cooling factor $\beta_\text{cool}$, the disc to star mass ratio and the Stokes number. We found that the dust to gas ratio increases with the cooling factor and decreases with the Stokes number, and that the relative temperature increases with the Stokes number and decreases with the cooling factor and the disc to star mass ratio. It is possible to explain these relationships by considering the interaction between dust particles and gas spiral arms. We compared our findings with \cite{boothclarke16} and found good agreement.
    \item We investigated the role of dust in gravitational instability, and found that the most unstable regions of the disc are the spiral arms, where the instability tends to be dust driven. In addition, we studied the relationship between the theoretical Jeans mass and the Stokes number. We found that the Jeans mass --- when the instability is dust driven --- can reach values of the order of the Earth mass.
    \item We observe three cases of dust collapse in our set of simulation, which occur (as expected) in the high disc mass models. The values of the clump masses obtained numerically are close to those predicted by linear theory. 
\end{enumerate}    

In applying our results to the possibility of early planet formation in Class 0/I discs, the main prerequisite is the requirement that dust is able to grow via coagulation to a large enough Stokes number, with a top-heavy particle mass function, in a short enough time. Our simulations that explicitly exhibited dust collapse had solid particles with an average Stokes number $\langle {\rm St} \rangle = 16$, which would correspond (rescaling our simulations to a disc size of $R_\text{out}=250$au) to particles with sizes between a few cm and a few metres. Fragmentation and radial drift pose barriers to growth to the required sizes \citep{Birnstiel10}, and further work is needed to assess whether there are circumstances where the required Stokes numbers can be reached. Simulations with a constant Stokes number, along with runs with ${\rm St} \sim 1$ (a regime which is numerically difficult to access using our code), would also help to better define the regime where dust can fragment in a gas disc that is itself stable against fragmentation.

In an evolutionary context, our results imply that planet formation --- if it is able to occur via the mechanism of dust-dominated gravitational disc instability --- is likely to occur toward the end of the self-gravitating phase. This is when the competing effects of particle trapping and particle excitation are jointly most favourable for collapse. The masses and orbital radii of the planets formed via dust collapse are qualitatively in agreement with those inferred for the HL~Tau system, and we speculate that they may form from the collapse of solids in the spiral arms of a formerly self-gravitating protostellar disc.

\section*{Acknowledgements}
The authors thank the referee for useful comments and suggestions that significantly improved the quality of the work.
In this work we used \textsc{splash} to create figures of the hydrodynamical simulations \citep{splash}. This project and the authors have received funding from the European Union’s Horizon 2020 research and innovation programme under the Marie Skłodowska-Curie grant agreement N. 823823 (DUSTBUSTERS RISE project). CL acknowledges support from Fulbright Commission through VRS scholarship. PJA acknowledges support from NASA TCAN award 80NSSC19K0639, and from award 644616 from the Simons Foundation. DJP acknowledges Australian Research Council funding via DP220103767 and DP180104235. The authors thank Sahl Rowther, Benedetta Veronesi, Cathie Clarke and Richard Booth for useful discussions. We thank Stony Brook Research Computing and  Cyberinfrastructure, and the Institute for Advanced Computational Science at Stony Brook University for access to the SeaWulf computing system, supported by National Science Foundation grant \# 1531492.

\section*{Data Availability}
The data presented in this article will be shared on reasonable request to the corresponding author.



\bibliographystyle{mnras}
\bibliography{example} 




\appendix

\section{Value of $\Lambda$}\label{applambda}
\begin{figure}
	\includegraphics[scale=0.425]{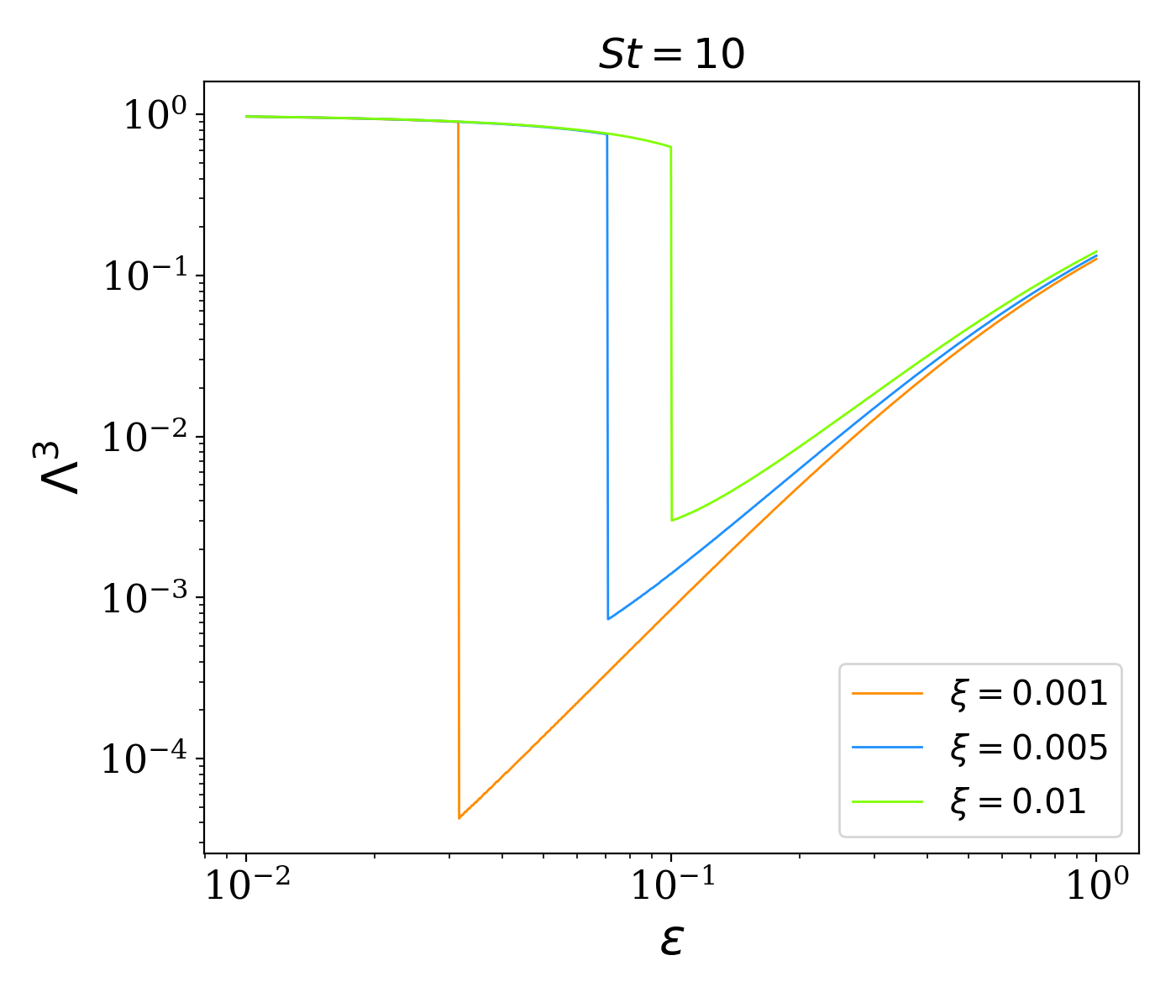}

 \includegraphics[scale=0.425]{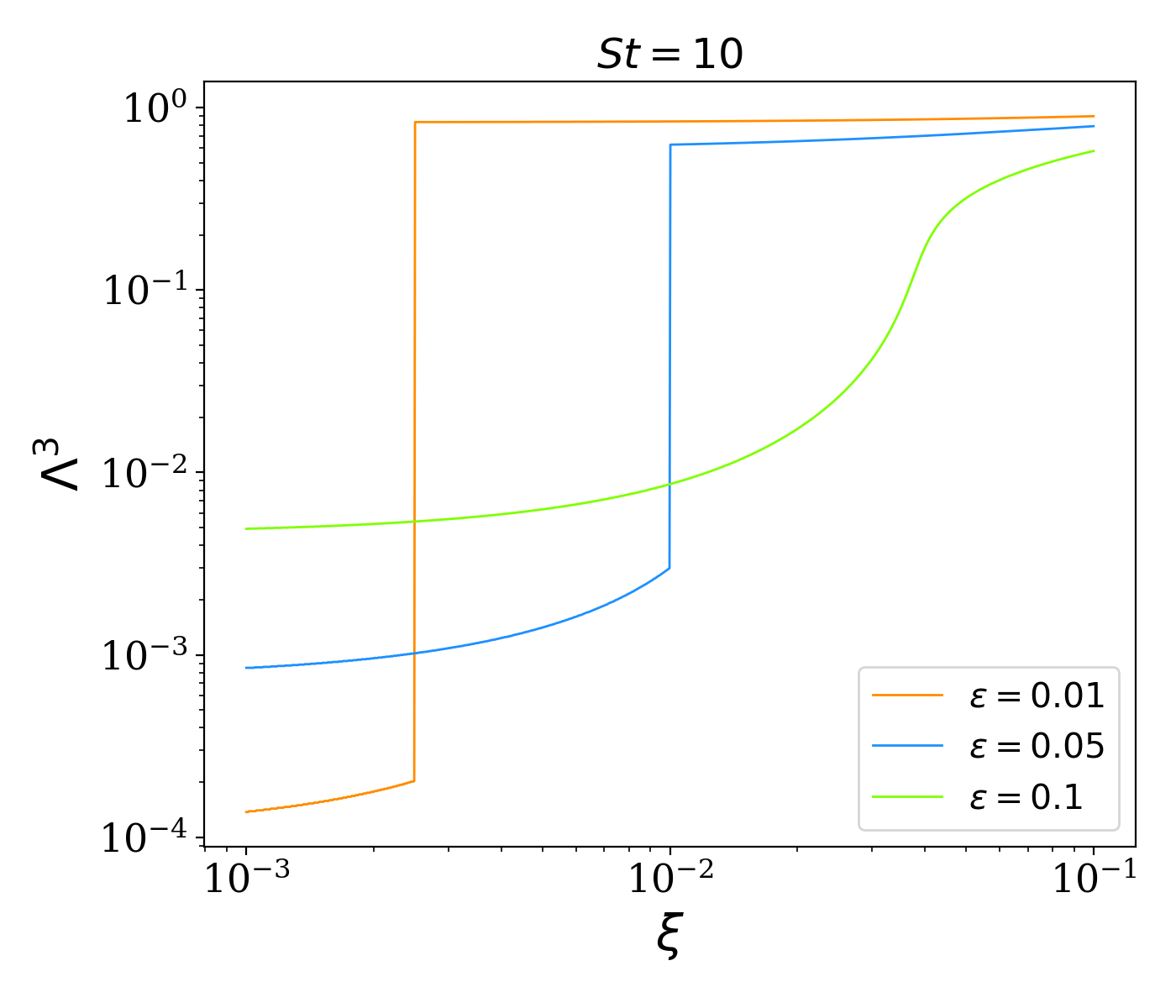}

 \includegraphics[scale=0.425]{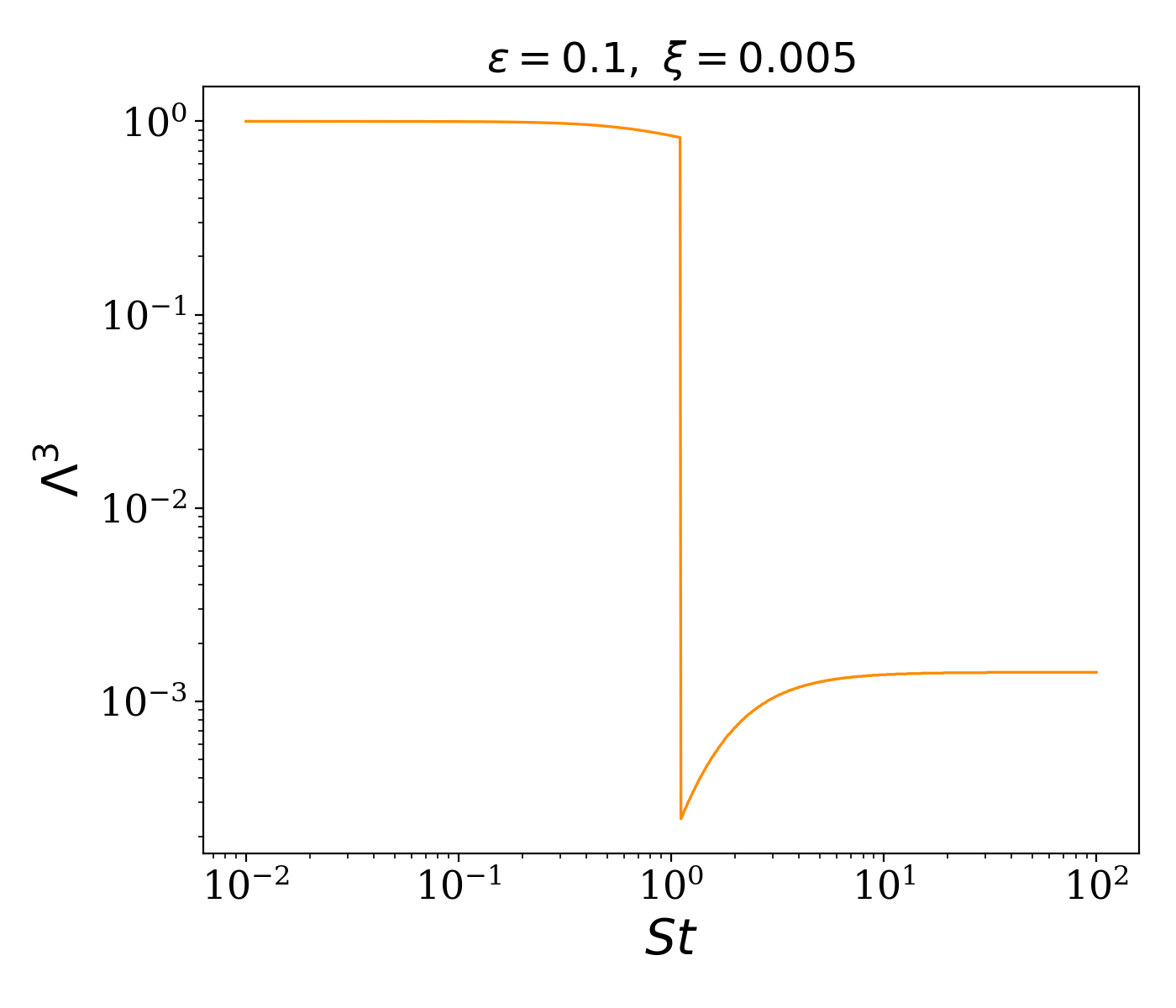}
    \caption{$\Lambda^3$ as a function of $(\epsilon, \xi, St)$, from top to bottom panel.}
    \label{lambda3}
\end{figure}

In this appendix we discuss the value of $\Lambda$ used to compute the Jeans mass for the two fluid components model. The value of $\Lambda$ can be obtained through the dispersion relation of the two fluid component model with drag force \citep{longa2fl}, by computing the most unstable wavelength having fixed $(\epsilon, \xi, St)$. Figure~\ref{lambda3} shows the value of $\Lambda^3$ as a function of $(\epsilon, \xi, St)$. The curves show a sharp fall: this is the point where instability becomes dust driven. When the instability is gas driven, the value of $\Lambda^3$ decreases with the dust to gas ratio. Then, it starts increasing since the mass of the dust becomes higher and higher, significantly contributing to the total density. Conversely, the value of $\Lambda^3$ increases with the relative temperature both in gas and dust driven regimes. Finally, $\Lambda^3$ decreases with the Stokes number.

\section{Resolution requirements}
\begin{figure}
	\includegraphics[scale=0.425]{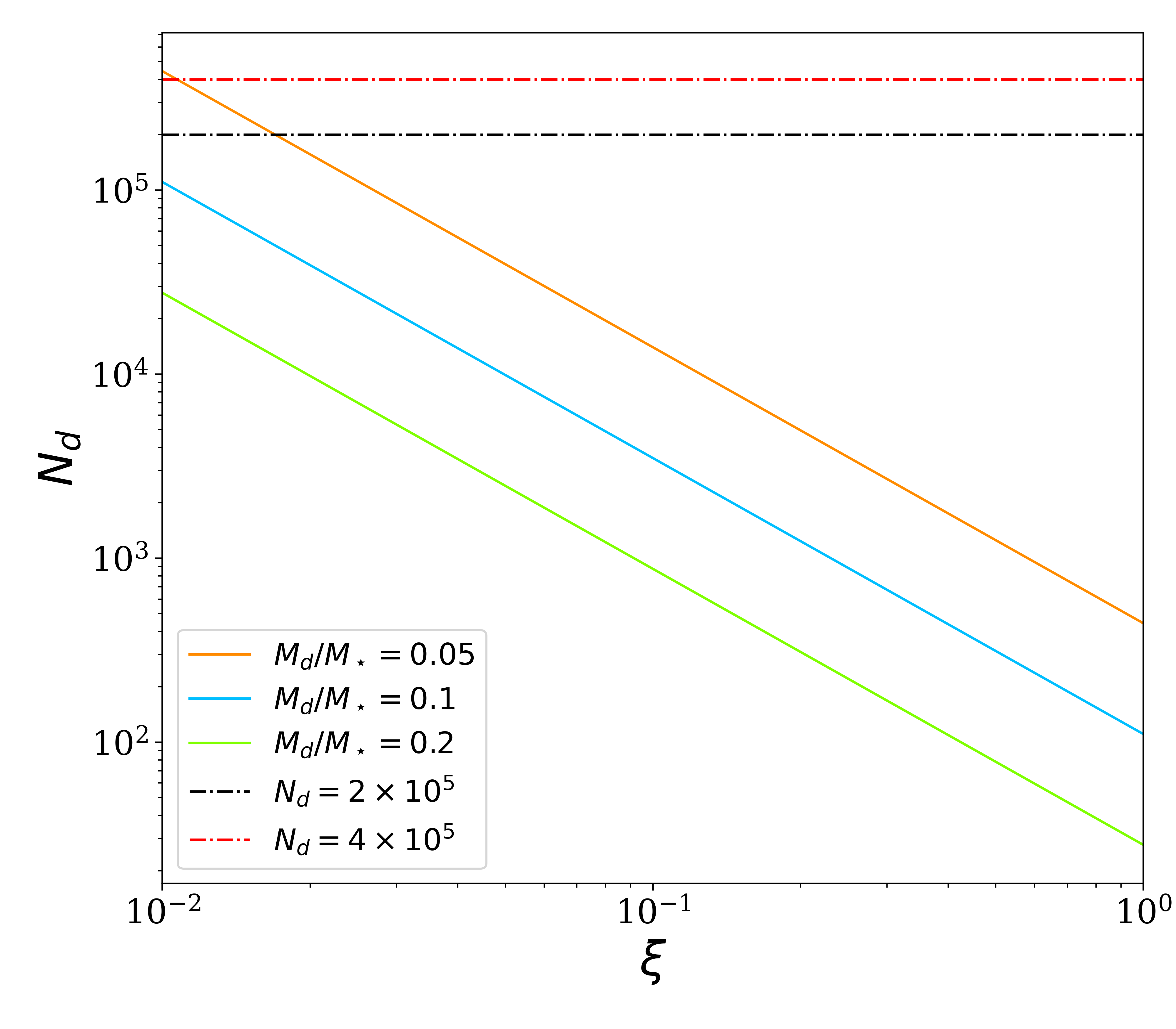}
    \caption{Minimum number of dust particles as a function of the relative temperature $\xi$ for different values of the disc-to-star mass ratio. The black line represents the value of this work}
    \label{resolution_requirement}
\end{figure}

In this appendix, we compute the minimum number of dust particles to resolve the expected Jeans mass: to study this problem, we refer to \cite{lodclarke}. In order to resolve the Jeans length, we require the smoothing length to be smaller than the disc height: so, the condition for dust particles is 
\begin{equation}\label{res_req}
   \frac{h_d}{H_d}<1.
\end{equation} 
Dust smoothing length and height can be written as a function of gas properties as follows
\begin{equation}
    H_d = H_g\xi^{1/2},
\end{equation}
\begin{equation}
    h_d = h_g \left(\frac{1}{100}\frac{N_g\rho_g}{N_d\rho_d}\right)^{1/3},
\end{equation}
where we assumed $M_\text{dust} / M_\text{gas} = 1/100$. By considering the gas disc to be marginally unstable ($Q_g=1$), condition~\ref{res_req} becomes 
\begin{equation}
    \frac{N_d}{2.5\times10^5} > \xi^{-3/2} \left(\frac{\epsilon}{0.01}\right)^{-1} \left(\frac{m(r)}{5\times10^{-3}}\right)^{-3} \left(\frac{M_d/M\star}{0.1}\right),
\end{equation}
where $\epsilon$ is the dust to gas ratio and $m(r)=\Sigma_g r^2 /M_\star$. If we write the disc mass as $M_d = \pi r_\text{out}^2 \Sigma_g$, $m(r) \simeq 1/\pi (M_d/M_\star)$, thus the resolution requirement is
\begin{equation}
    N_d > 4\times10^4 \left(\frac{\xi}{0.1}\right)^{-3/2}\left(\frac{\epsilon}{0.01}\right)^{-1}\left(\frac{M_d/M_\star}{0.1}\right)^{-2}.
\end{equation}
Figure~\ref{resolution_requirement} shows the minimum number of dust particles required to resolve the Jeans length as a function of the relative temperature $\xi$ for $M_d/M_\star=\{0.05,0.1,0.2\}$, for a fixed dust to gas ratio $\epsilon=0.1$: even for extreme values of $\xi$, the standard resolution we have chosen ($N_d=2\times10^5$) is sufficient to resolve the Jeans length. One should note that we never reach $\xi\sim10^{-2}$, and in addition we have underestimated the dust to gas ratio: indeed, inside spiral arms it could reach values $>0.1$, making our $N_d$ choice even safer.

\section{Large and small dust particles comparison}\label{appb}

\begin{figure*}
	\includegraphics[width=\textwidth]{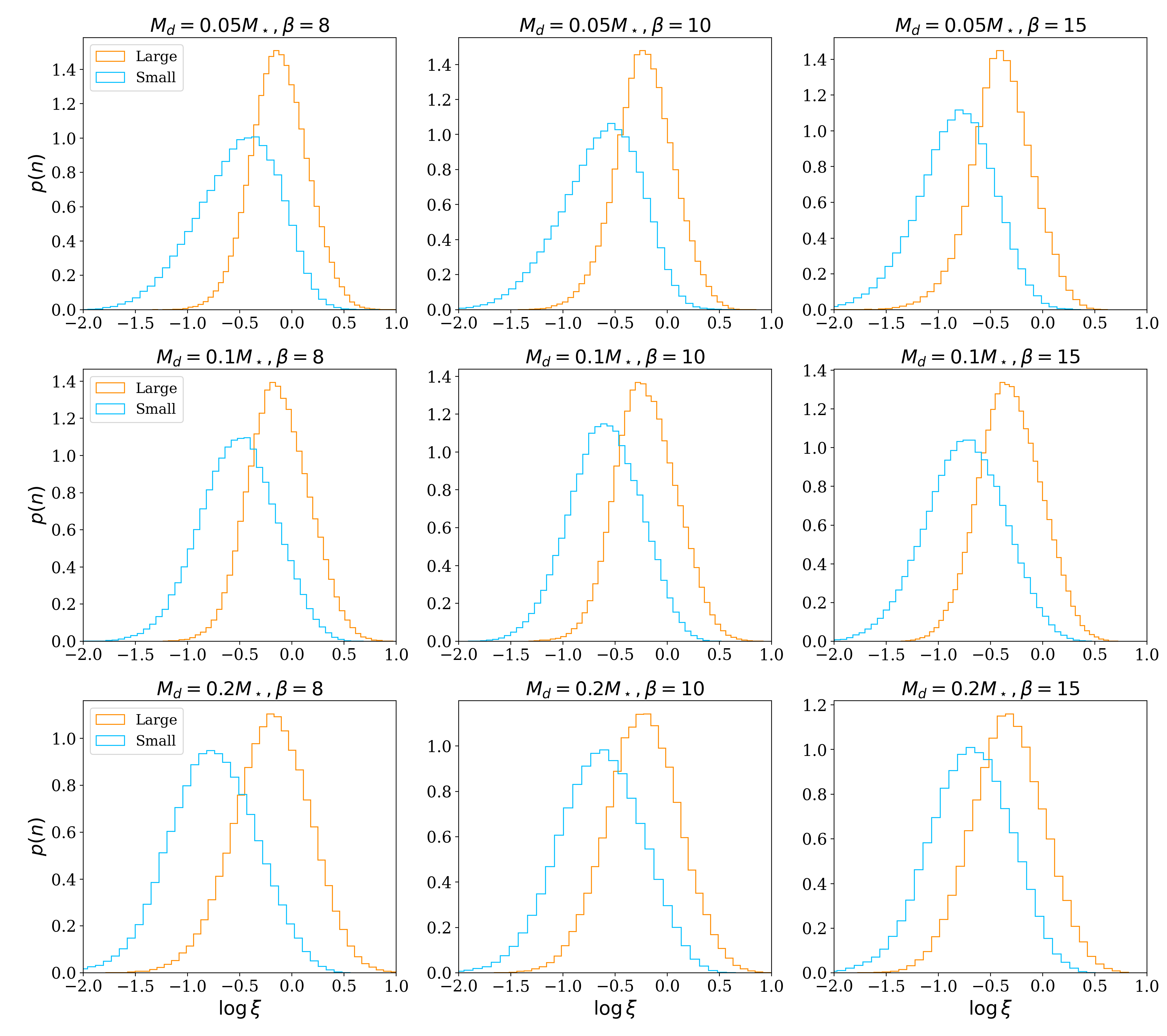}
    \caption{Comparison between large and small dust particles.}
    \label{h3}
\end{figure*}

Figure~\ref{h3} show the complete set of histograms of the dust to gas ratio and the relative temperature, for small and large dust particles.

\begin{figure*}
	\includegraphics[width=\textwidth]{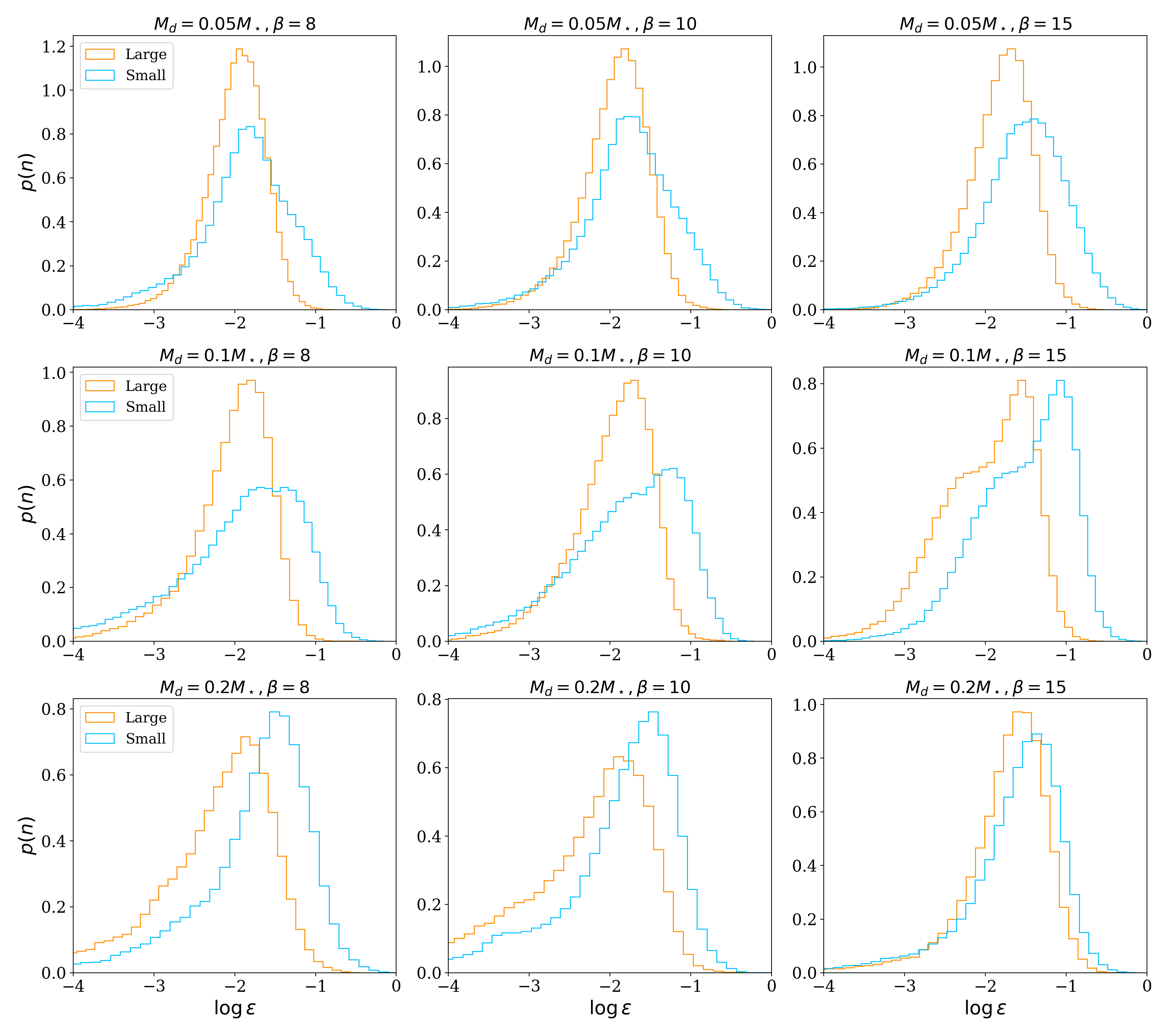}
     \caption{Comparison between large and small dust particles.}
\end{figure*}

\bsp	
\label{lastpage}
\end{document}